\documentclass[journal,twoside,web]{ieeecolor2}
\usepackage{generic}
\usepackage{amssymb}
\usepackage{hyperref}
\usepackage{amsmath,bm}
\usepackage{algorithm}
\usepackage{algpseudocode}
\usepackage{array}
\usepackage{subfig}
\usepackage{cite}
\usepackage{tabularx}
\usepackage{tabulary}
\usepackage{mathrsfs}
\usepackage{graphicx}
\usepackage{epstopdf}
\usepackage{multirow}
\usepackage{makecell}
\usepackage{booktabs}
\usepackage{url}
\usepackage[dvipsnames]{xcolor}
\hypersetup{hidelinks}
\definecolor{IEEEBlue}{RGB}{0,138,218}
\def\BibTeX{{\rm B\kern-.05em{\sc i\kern-.025em b}\kern-.08em
T\kern-.1667em\lower.7ex\hbox{E}\kern-.125emX}}
\begin{document}
\bstctlcite{IEEEexample:BSTcontrol}
\title{Robust Sparse Bayesian Learning\\Based on Minimum Error Entropy for Noisy\\High-Dimensional Brain Activity Decoding}
\author{Yuanhao Li, Badong Chen, \IEEEmembership{Senior Member, IEEE,} Wenjun Bai, Yasuharu Koike, and Okito Yamashita
\thanks{This work was supported in part by Innovative Science and Technolo- gy Initiative for Security under Grant JPJ004596 ATLA, in part by Moon- shot Program 9 under Grant JPMJMS2291, and in part by National Natural Science Foundation of China under Grant 62436005. \emph{(Corresponding author: Yuanhao Li.)}}
\thanks{Yuanhao Li and Okito Yamashita are with the Center for Advanced Int- elligence Project, RIKEN, Tokyo 103-0027, Japan, and also with the Department of Computational Brain Imaging, ATR Neural Information Analysis Laboratories, Kyoto 619-0237, Japan. (e-mail: yuanhao.li@riken.jp)}
\thanks{Badong Chen is with the Institute of Artificial Intelligence and Roboti- cs (IAIR), Xi'an Jiaotong University, Xi'an 710049, China.}
\thanks{Wenjun Bai is with the Department of Computational Brain Imaging, ATR Neural Information Analysis Laboratories, Kyoto 619-0237, Japan.}
\thanks{Yasuharu Koike is with the Institute of Integrated Research (IIR), Instit- ute of Science Tokyo, Yokohama 226-8501, Japan.}
\thanks{The MATLAB code is available at \href{https://github.com/liyuanhao-git/SBL-MEE}{\textcolor{IEEEBlue}{\textit{github.com/liyuanhao-git/SBL-MEE}}}.}}

\maketitle

\begin{abstract}
\textit{Objective}: Sparse Bayesian learning provides an effective framework to solve high-dimensional problems in brain signal decoding. However, conventional likelihoods regarding data distributions, such as Gaussian or Bernoulli, are potentially inadequate for handling the noisy recordings of brain activity. Hence, this work aims to formulate a robust sparse Bayesian learning framework to address noisy high-dimensional brain activity decoding. \textit{Methods}: Motivated by the commendable robustness of the minimum error entropy learning criterion for addressing non-Gaussian signals, this study reformulated the sparse Bayesian learning framework under a generalized Bayesian paradigm, in which the model parameter is regulated with the minimum error entropy loss rather than a conventional likelihood function. \textit{Results}: Our developed SBL-MEE algorithm was evaluated with two real-world brain decoding tasks of regression and classification scenarios, respectively. Experimental results demonstrated that our approach not only realizes superior brain decoding performance than existing methods, but also presents more physiologically interpretable decoder patterns. \textit{Conclusion}: Although minimum error entropy is not constructed from an arbitrary probabilistic distribution, it is effective to establish noise-robust inference in sparse Bayesian learning method. \textit{Significance}: This work provides a powerful tool to improve brain activity decoding capability, particularly regarding the noisy high-dimensional setting, thus promoting biomedical engineering applications such as brain-computer interface.

\end{abstract}


\section{Introduction}
\label{sec:introduction}

\IEEEPARstart{D}{ecoding} high-level cognitive intentions and perceptual states from brain activity recording has promoted various successful applications of brain-computer interfaces (BCI) \cite{wolpaw2000brain,edelman2025non} and promising neuroscience investigations \cite{haxby2014decoding,rybavr2022neural,robinson2023visual}. Despite these advances achieved by various machine learning methods, brain activity decoding has been consistently challenged by the following two obstacles. First, brain activity recordings usually exhibit high-dimensional feature space, encompassing copious voxels for the functional magnetic resonance imaging (fMRI), and multiple channels with high temporal resolution leveraged in electroencephalogram (EEG) or electrocorticogram (ECoG). However, the number of labeled training samples is commonly limited in the practical neural decoding task. This is not merely due to the high cost and long duration of neural data collection, but also because usable samples depend on carefully controlled experimental paradigms and synchronization between stimulus and neural response. Moreover, differences in the experimental protocol prevent data of different works from being aggregated into a large training set. This results in a high-dimensional low-sample-size problem, in which the traditional machine learning methods such as ordinary least square (OLS) regression would suffer significant overfitting with poor generalization on testing samples \cite{van2012brain,tang2021dimensionality}. Second, neural recording signals are typically degraded by a complex mixture of different noise components. For example, electromagnetic neural signal including EEG and ECoG is prone to environmental noises, system-related noises, and physiological artifacts \cite{ball2009signal}, while fMRI recording is usually corrupted by physiological noises and cephalic motion artifacts \cite{liu2016noise}. The mixture of these noise components leads to a complex noise distribution for brain recording signal, which is typically non-Gaussian and highly variable across sessions and subjects. As a result, the training of decoding models using conventional machine learning approaches will be significantly deteriorated, leading to poor learning performance from neural activity data. These two problems highlight the necessity for developing new machine learning approach that can solve the high-dimensional nature and data noise concurrently for achieving accurate brain decoding.

To alleviate the high-dimensional problem in decoding brain activities with small training datasets, sparse Bayesian learning (SBL) has emerged as an adequate framework which can prune automatically the less relevant features by a Bayesian inference paradigm \cite{faul2001analysis,tipping2001sparse}. Compared to the dimensionality reduction techniques, e.g., principal component analysis (PCA), SBL can provide a superior interpretability through using a subset of the original covariates with feature selection. In addition, different from the sparsity-promoting $L_1$-regularization that necessitates manual adjustments on the model sparsity, SBL enables a self-propelled model sparsity control, which is easier to implement for the real-world neural decoding scenarios. These advantages have contributed to the widespread practice of SBL in different brain activity decoding tasks, which can mainly be categorized into regression task \cite{ganesh2008sparse,toda2011reconstruction,yoshimura2012reconstruction,umeda2019decoding,wang2023sparse} and classification task \cite{miyawaki2008visual,shibata2011perceptual,yahata2016small,horikawa2017generic,ganesh2018utilizing,shi2021galvanic}. However, these existing applications of SBL in brain decoding have not fully considered the complex noise distribution which usually utilize conventional data assumptions such as Gaussian or Bernoulli. Therefore, the corresponding likelihood functions exhibit potentially insufficient robustness in dealing with noisy brain data, and thus SBL may suffer performance deterioration practically.

On the other hand, to solve the inherent measurement noises in brain activity recordings, various machine learning methods have been developed from different perspectives. For example, denoising techniques, such as independent component analysis (ICA) based artifact exclusion \cite{jung2000removing,escudero2007artifact,hamaneh2013automated}, have been effectively employed to ameliorate the signal quality of neural recordings. However, it is difficult to guarantee that all the recording noises can be totally removed. Another pathway to solve this problem is to develop robust objective function for the machine learning model that enables correct model training with a noisy dataset. Notably, the information theoretic learning (ITL) \cite{principe2010information} provides an efficient framework to develop the robust objective function for different machine learning tasks. In particular, two learning criteria in ITL have been attracting considerable attention from the community, named maximum correntropy criterion (MCC) \cite{liu2007correntropy} and minimum error entropy (MEE) \cite{erdogmus2002error}. MCC is adequate for addressing the outlier and the extremely heavy-tailed noise, while MEE demonstrates a superior flexibility that is moreover well-suited for multimodal and moderately heavy-tailed noises \cite{chen2016insights,chen2019effects}. MCC and MEE have both been leveraged to develop robust brain decoding algorithms \cite{chen2018common,zheng2020broad,zheng2020mixture,li2021restricted,li2023partial,zheng2023quantized}. Nevertheless, few of these advances can be directly adopted for high-dimensional brain decoding tasks, since they basically lack explicit sparsity control, undergoing serious overfitting in the high-dimensional scenario.

To realize superior brain decoding performance with solving the two problems of high-dimensional and noisy neural signals simultaneously, the purpose of this study is to propose a robust SBL framework that can reduce the effects of recording noises in brain decoding. The methodological intuition is that sparsity promotion and noise robustness should be achieved at different levels of a Bayesian model. Sparse priors are effective to select informative features, whereas robustness against noisy samples is essentially determined by the likelihood function in the data-fitting aspect. Thus, improving the robustness is more naturally achieved by the likelihood. From this perspective, our previous works have proposed an MCC-based robust likelihood function to construct a sparse Bayesian correntropy learning framework called SBCL, which showed promising performance in various brain signal analysis tasks \cite{li2023adaptive,li2023correntropy,li2025sparse,li2025correntropy}. Compared to MCC, MEE is more flexible in characterizing complex data distribution and is less restricted by simple parametric assumptions. Employing MEE to formulate SBL method hence exhibits better potentials to handle heterogeneous brain data. Motivated by this intuition, in the present study, we proposed a new robust SBL framework based on MEE. The main contributions are outlined as follows:

\begin{itemize}
\item[1.]
We proposed a new robust SBL approach, named as SBL-MEE, under a generalized Bayesian framework, for which the MEE criterion is utilized as the optimization target for the posterior updating, leading to a unified framework that is applicable to both regression and classification settings.

\item[2.]
To effectively estimate the model parameter for the MEE-based generalized posterior distribution, this work derived a tractable learning algorithm by variational inference and Laplacian approximation.

\item[3.]
The proposed SBL-MEE approach was evaluated through two real-world neural activity decoding tasks, considering regression-based and classification-based context settings, respectively.

\item[4.]
The experimental results demonstrated that our developed SBL-MEE framework not only realizes superior decoding performance for brain data compared to existing methods, but also exhibits a more physiologically plausible decoder pattern.
\end{itemize}

We organize the remainder of this paper using the following structure. Section \ref{sec:method} introduces the conventional SBL approach and the robust MEE learning criterion first, and then elaborates the reformulated SBL framework based on MEE. Then Section \ref{sec:results} describes two real-world brain decoding scenarios to assess the proposed approach, and illustrates the experimental results, with performance comparisons to baselines and state-of-the-art methods. Afterwards, in Section \ref{sec:discussion} we also provide discussion regarding the proposed method. Finally this paper is concluded in Section \ref{sec:conclusion}. The code to reproduce the results in this research is available at GitHub repository \href{https://github.com/liyuanhao-git/SBL-MEE}{\textcolor{IEEEBlue}{\textit{https://github.com/liyuanhao-git/SBL-MEE}}}. Appendix A presents a comprehensive literature review on previous studies that are related to the present paper, elaborating their distinction from this work (see supplementary material).

\section{Method}
\label{sec:method}

\subsection{Sparse Bayesian Learning}

To facilitate the formulation of SBL using a unified skeleton which includes regression and classification tasks concurrently, we leverage the generalized linear model (GLM) \cite{nelder1972generalized} to define the problem settings. For each input $\textbf{x}=\left[x_1,\cdots,x_D\right]^\top\in\mathbb{R}^D$ that represents a $D$-dimensional vector, GLM employs the link function with the following expression to establish the relation between the input covariate $\textbf{x}$ and the desired response variable
\begin{equation}
\label{Eq.Glm1}
\mathbb{E}\left[t|\textbf{x}\right]=g^{-1}(\textbf{x}^\top\textbf{w})
\end{equation}
where $\textbf{w}=\left[w_1,\cdots,w_D\right]^\top\in\mathbb{R}^D$ is the model parameter, and $\mathbb{E}\left[t|\textbf{x}\right]$ indicates the expectation of desired output $t$ conditioned on $\textbf{x}$. For the linear regression, the identical mapping is utilized as the link function, i.e., $\mathbb{E}\left[t|\textbf{x}\right]=\textbf{x}^\top\textbf{w}$. For logistic regression, to classify the categorical response $t\in\left\{0,1\right\}$, the link function employs the sigmoid formula $\mathbb{E}\left[t|\textbf{x}\right]=1/\left(1+\exp\left(-\textbf{x}^\top\textbf{w}\right)\right)$.

In order to obtain the optimal model parameter, a probability distribution is assumed for the response variable. For example, in linear regression tasks, we leverage the Gaussian assumption
\begin{equation}
\label{Eq.GaussianModel}
p(t|g^{-1}(\textbf{x}^\top\textbf{w}))=\mathcal{N}(t|\textbf{x}^\top\textbf{w},\sigma^2)
\end{equation}
which represents a Gaussian distribution with mean value $\textbf{x}^\top\textbf{w}$ and variance $\sigma^2$. On the other hand, for logistic regression, the response variable is assumed to obey the Bernoulli distribution
\begin{equation}
\label{Eq.BinomialModel}
\begin{split}
p(t=1|g^{-1}(\textbf{x}^\top\textbf{w}))=&\frac{1}{1+\exp(-\textbf{x}^\top\textbf{w})}\\
p(t=0|g^{-1}(\textbf{x}^\top\textbf{w}))=&\frac{\exp(-\textbf{x}^\top\textbf{w})}{1+\exp(-\textbf{x}^\top\textbf{w})}\\
\end{split}
\end{equation}
In practice, given a finite dataset $\left\{\left(\textbf{x}_i,t_i\right)\right\}_{i=1}^N$ with $N$ samples, the likelihood function could be written with assuming sample independence
\begin{equation}
\label{Eq.Likelihood}
p(\textbf{t}|\textbf{w})=\prod_{i=1}^{N}p(t_i|g^{-1}(\textbf{x}_i^\top\textbf{w}))
\end{equation}
in which $\textbf{t}$ denotes the whole dataset. The maximum likelihood estimation (MLE) of the model parameter can be thus obtained by maximizing the logarithmic form of the likelihood function
\begin{equation}
\label{Eq.LikelihoodMax}
\begin{split}
\textbf{w}_{MLE}&=\arg\underset{\textbf{w}\in\mathbb{R}^D}{\max}\log p(\textbf{t}|\textbf{w})\\
&=\arg\underset{\textbf{w}\in\mathbb{R}^D}{\max}\sum_{i=1}^{N}\log p(t_i|g^{-1}(\textbf{x}_i^\top\textbf{w}))\\
\end{split}
\end{equation}
Notably, with the conventional Shannon's definition of entropy, the negative log-likelihood in (\ref{Eq.LikelihoodMax}) exhibits an empirical estimate of the cross entropy quantity between the true data distribution and the learned distribution. When substituting in the Gaussian and Bernoulli forms, one could observe that, the MLE solution of linear regression with the Gaussian assumption is equivalent to utilizing mean squared error (MSE) loss, while the Bernoulli likelihood in logistic regression equals the binary cross entropy loss.

For the high-dimensional problem in which one has $D>N$, the MLE will lead to serious overfitting on the training dataset. To alleviate this persistent problem, the SBL framework offers a powerful approach that infers the relevance of each covariate and thus removes less important dimensions. For each element of the model parameter, SBL uses a Gaussian prior assumption
\begin{equation}
p(w_d|a_d)=\mathcal{N}(w_d|0,a_d^{-1})
\end{equation}
in which the inverse variance $a_d$ is named relevance parameter, where a large value of $a_d$ implies that the corresponding model parameter $w_d$ is tightly distributed at zero, therefore exhibiting low relevance. Thus, the prior distribution for the whole model parameter is
\begin{equation}
\label{Eq.Prior1}
p(\textbf{w}|\textbf{a})=\prod_{d=1}^{D}p(w_d|a_d)=\prod_{d=1}^{D}\mathcal{N}(w_d|0,a_d^{-1})
\end{equation}
In addition, to facilitate a fully Bayesian inference framework, one can further leverage the following non-informative Jeffreys prior distribution \cite{gelman1995bayesian} on each entry of the relevance parameter
\begin{equation}
\label{Eq.Prior2}
p(\textbf{a})=\prod_{d=1}^{D}p(a_d)=\prod_{d=1}^{D}a_d^{-1}
\end{equation}
Then the joint posterior distribution regarding model parameter and relevance parameter is computed by the following formula
\begin{equation}
p(\textbf{w},\textbf{a}|\textbf{t})=\frac{p(\textbf{t}|\textbf{w})p(\textbf{w}|\textbf{a})p(\textbf{a})}{p(\textbf{t})}
\end{equation}
where the integral $p(\textbf{t})=\int p(\textbf{t}|\textbf{w})p(\textbf{w}|\textbf{a})p(\textbf{a})d\textbf{a}d\textbf{w}$ is difficult to acquire analytically. To address this obstacle, the variational inference technique \cite{blei2017variational} provides an effective way to calculate the maximum a posteriori (MAP) estimation or posterior mean of model parameter and relevance parameter. During the model training, the elements in $\textbf{a}$ that correspond to irrelevant features will become arbitrarily large, indicating a compact distribution around zero regarding the model parameter \cite{wipf2007new}. This selection process of relevant feature is known as the automatic relevance determination (ARD) mechanism, which serves as the essential component of SBL. In practice, a certain feature will be pruned if its relevance parameter exceeds the predetermined threshold. Compared to the Laplace prior distribution adopted by LASSO \cite{tibshirani1996regression}, the hierarchical Gaussian prior with a Jeffreys hyper-prior provides two advantages. First, it avoids manually determining a sparsity-controlling parameter since the relevance parameters can be automatically inferred from data. Second, the equivalent marginal prior exhibits heavier tail, hence alleviating excessive shrinkage of large weights and strongly encouraging parameter pruning \cite{wipf2007new}.

\subsection{Minimum Error Entropy}
\label{Sec.Mee}

To realize robust model learning, the MEE learning criterion has been developed as a competent substitute for the traditional optimization objectives \cite{erdogmus2002error,chen2016insights,chen2018common,li2021restricted,zheng2023quantized,chen2018quantized}, which can capture the higher-order statistical information of residuals by minimizing the entropy of the difference between prediction and desired output. To estimate the entropy for prediction error $e=t-\hat{t}$, in which $\hat{t}$ represents the current model output, MEE leverages the $\alpha$-order Renyi’s entropy defined by the following equation \cite{principe2010information,erdogmus2002error}
\begin{equation}
H_{\alpha}(e)=\frac{1}{1-\alpha}\log\int\left[p(e)\right]^{\alpha}de
\end{equation}
in which $p(e)$ represents the probability density function (PDF) of residuals. Commonly, MEE adopts $\alpha=2$ for computational simplicity, which thus leads to the following objective function
\begin{equation}
\begin{split}
\textbf{w}_{MEE}&=\arg\underset{\textbf{w}}{\min}-\log\int\left[p(e)\right]^2de\\
&=\arg\underset{\textbf{w}}{\max}\int\left[p(e)\right]^2de\\
\end{split}
\end{equation}

\noindent where the second equation is derived as the logarithm function is a monotonically increasing function. Therefore, the learning target for MEE can be regarded as maximizing the expectation value of error PDF $\mathbb{E}\left[p(e)\right]=\int\left[p(e)\right]^2de$. In practice, one can utilize a finite dataset $\left\{e_i\right\}_{i=1}^N$ to acquire the empirical estimate for $\mathbb{E}\left[p(e)\right]$ by adopting the nonparametric PDF estimator \cite{parzen1962estimation}, yielding
\begin{equation}
\label{Eq.Mee}
\begin{split}
\textbf{w}_{MEE}&=\arg\underset{\textbf{w}}{\max}\mathbb{E}\left[p(e)\right]=\arg\underset{\textbf{w}}{\max}\frac{1}{N}\sum_{i=1}^{N}\hat{p}\left(e_i\right)\\
&=\arg\underset{\textbf{w}}{\max}\frac{1}{N}\sum_{i=1}^{N}\frac{1}{N}\sum_{j=1}^{N}k_{\sigma}\left(e_i-e_j\right)\\
&=\arg\underset{\textbf{w}}{\max}\frac{1}{N^2}\sum_{i=1}^{N}\sum_{j=1}^{N}k_{\sigma}\left(e_i-e_j\right)\\
\end{split}
\end{equation}
where $\hat{p}\left(e_i\right)=\frac{1}{N}\sum_{j=1}^{N}k_{\sigma}\left(e_i-e_j\right)$ is the non-parametrically estimated error PDF at $e_i$, and $k_{\sigma}\left(x\right)=\exp\left(-\frac{x^2}{2\sigma^2}\right)$ indicates Gaussian kernel function with bandwidth $\sigma$. The normalization constant is omitted, since it does not influence the optimization with respect to $\textbf{w}$. Note that, the Gaussian kernel function aims to obtain a nonparametric estimate of error PDF via the Parzen window method \cite{parzen1962estimation}, rather than introducing a distance metric. This learning criterion is in particular applicable for addressing complex residual distribution, because it does not impose strict distributional assumptions. As a result, MEE can accommodate non-Gaussian perturbation in brain recording data, and provide superior decoding capacity \cite{chen2018common,li2021restricted,zheng2023quantized}. Compared to MCC \cite{liu2007correntropy}, MEE concentrates on the uncertainty of error distribution instead of emphasizing only local similarity around zero. Thus, MEE is more flexible for modeling a complex data distribution \cite{chen2016insights,chen2019effects}. Appendix B presents a detailed comparison between MEE and MCC concerning their properties (see supplementary material).

\subsubsection{MEE-Based Regression}

Although the objective function of original MEE (\ref{Eq.Mee}) is effective for dealing with various noise distributions in the regression task, it is limited by a substantial computational demand that results from the double summation in (\ref{Eq.Mee}). To this end, a computationally efficient variant of MEE was proposed by estimating the error PDF using a quantization approach, called as quantized MEE (QMEE) \cite{chen2018quantized}. Specifically, QMEE constructs a quantization codebook $C=\left\{c_1,\cdots,c_M\right\}$ containing $M$ elements ($M\ll N$) to represent the whole error set. Each error sample is mapped to a specific element $c_j$ using a clustering-based method, and $\eta_j$ denotes the number of error samples which are quantized to $c_j$. Thus, the objective function of QMEE is
\begin{equation}
\label{Eq.Qmee}
\begin{split}
\textbf{w}_{QMEE}&=\arg\underset{\textbf{w}}{\max}\frac{1}{N^2}\sum_{i=1}^{N}\sum_{j=1}^{N}k_{\sigma}\left(e_i-Q\left[e_j\right]\right)\\
&=\arg\underset{\textbf{w}}{\max}\frac{1}{N^2}\sum_{i=1}^{N}\sum_{j=1}^{M}\eta_j\cdot k_{\sigma}\left(e_i-c_j\right)\\
\end{split}
\end{equation}
in which $Q\left[\cdot\right]$ is a quantization operator which clusters $\eta_j$ error samples to the quantization element $c_j$. Clearly, one can know that $\sum_{j=1}^{M}\eta_j=N$. Consequently, the complexity is decreased from $\mathcal{O}(N^2)$ to $\mathcal{O}(MN)$ where $M\ll N$. Both theoretical and experimental results demonstrated that by formulating a proper codebook $C$, QMEE can achieve similar learning performance as the original MEE with evidently reducing the computational efforts \cite{chen2018quantized}. Algorithm \ref{Algo.Quantization} summarizes the steps for constructing the codebook.

\begin{algorithm}[h]
\caption{\textit{Quantization procedures}\cite{chen2018quantized}}
\label{Algo.Quantization}
\begin{algorithmic}[1]
\State \textbf{input}:

error dataset $\left\{e_i\right\}_{i=1}^{N}$

\State \textbf{initialize}:

quantization codebook $C=\left\{e_1\right\}$

\State \textbf{parameter setting}:

quantization threshold $\varepsilon$

\For{$i=2,\cdots,N$}
\State calculate the minimum distance between $e_i$ and all the elements in $C$ by $\min\left|e_i-C\left(j\right)\right|$, where $C\left(j\right)$ represents the $j$-th element in $C$
\If{$\min\left|e_i-C\left(j\right)\right|\leqslant\varepsilon$}
\State maintain the codebook unchanged and quantize $e_i$ to the nearest element, i.e. $Q\left[e_i\right]=C\left(j^*\right)$, in which $j^*=\arg\min_j\left| e_i-C\left(j\right) \right|$
\Else
\State update the codebook by $C=\left\{C,e_i\right\}$, and quantize $e_i$ through $Q\left[e_i\right]=e_i$
\EndIf
\EndFor
\State \textbf{output}:

quantization codebook $C=\left\{c_1,\cdots,c_M\right\}$
\end{algorithmic}
\end{algorithm}

\subsubsection{MEE-Based Classification}

Remarkably MEE and QMEE are not directly applicable to classification, because the optimal error distribution for binary logistic regression exhibits a three-modal characteristic located at -1, 0, and 1, resulting from false negatives, accurate prediction, and false positives, respectively \cite{li2021restricted,de2013minimum}. Unconstrained entropy minimization cannot impose explicitly this structure, which may yield suboptimal classifier, especially in noisy scenarios. To solve this limitation, restricted MEE (RMEE) was developed by introducing a fixed codebook $C=\left\{0,-1,1\right\}$ within the QMEE skeleton \cite{li2021restricted}. The restricted codebook qualifies the model optimization towards the optimal three-modal error distribution, which maximizes inner-product similarity between the current error PDF and the optimal form. The objective function of RMEE can be formulated as follows:
\begin{equation}
\label{Eq.Rmee}
\textbf{w}_{RMEE}=\arg\underset{\textbf{w}}{\max}\frac{1}{N^2}\sum_{i=1}^{N}{\left( \begin{array}{c}
\eta_0\cdot k _{\sigma}\left(e_i\right)\\
+\eta_{-1}\cdot k _{\sigma}\left(e_i+1\right)\\
+\eta_1\cdot k _{\sigma}\left(e_i-1\right)\\
\end{array} \right)}
\end{equation}
which is in essence a special case of QMEE with the codebook $C=\left\{0,-1,1\right\}$, accompanied by the quantization numbers $\eta_0$, $\eta_{-1}$, and $\eta_1$. The weighting coefficients can be estimated using a preliminary classifier by splitting all the training samples into correct predictions, false negatives, and false positives, through
\begin{equation}
\label{Eq.RmeeHyper}
\begin{split}
\eta_0&=\#\left[e\in\left(-0.5,0.5\right)\right]\\
\eta_{-1}&=\#\left[e\in\left(-1,-0.5\right)\right]\\
\eta_1&=\#\left[e\in\left(0.5,1\right)\right]\\
\end{split}
\end{equation}
where $\#\left[\cdot\right]$ indicates counting the relevant samples that satisfy the condition. The interval $e\in\left(-0.5,0.5\right)$ represents correctly classified samples, while errors less than $-0.5$ and greater than $0.5$ result from false negatives and false positives, respectively, as formulated in (\ref{Eq.RmeeHyper}) for determining the hyperparameter \cite{li2021restricted}. By estimating the quantization numbers from training samples with the empirical occurrence of each sample category, RMEE obtains an effective estimate for the weight, enabling the model learning towards the optimal three-mode error distribution with appropriate weights. This formulation preserves the robustness of MEE while extending its applicability to classification tasks, in particular for the noisy condition with considerable samples contaminated by erroneous labels and deviated attribute values.

\subsection{Robust Sparse Bayesian Learning via MEE}

To ameliorate the robustness of SBL for the real-world noisy high-dimensional brain decoding scenarios, we aim to propose a reformulated SBL approach by leveraging the MEE criterion. Recall that, the inadequate robustness of the conventional SBL framework results from the dependence on the overly idealized assumptions regarding data distributions, such as the Gaussian or Bernoulli models in (\ref{Eq.GaussianModel})(\ref{Eq.BinomialModel}), which are incorporated into SBL by the likelihood function in (\ref{Eq.Likelihood}). Therefore, the purpose of this study can be naturally devised as integrating the MEE learning criterion into the SBL skeleton, rather than depending on those conventional likelihood models. As introduced in Section \ref{Sec.Mee}, MEE objectives for regression and classification can be unified as:
\begin{equation}
\label{Eq.AllMee}
\underset{\textbf{w}}{\max}\;\mathcal{J}_{MEE}=\sum_{i=1}^{N}\sum_{j=1}^{M}\eta_j\cdot k_{\sigma}\left(e_i-c_j\right)
\end{equation}
despite the different configurations concerning the quantization element $c_j$ and weight $\eta_j$ in regression and classification tasks, respectively. The denominator $\frac{1}{N^2}$ is omitted as it is fixed given a finite dataset. Moreover, we can define the MEE-induced loss
\begin{equation}
\label{Eq.MeeLoss}
\mathcal{L}_{MEE}\triangleq-\mathcal{J}_{MEE}
\end{equation}
so that maximizing $\mathcal{J}_{MEE}$ is equivalent to minimizing the loss $\mathcal{L}_{MEE}$.

However, although this MEE-induced loss function provides a unified while robust learning criterion, it is not obtained from the negative log-likelihood of an arbitrary probabilistic density function. Particularly, one could perceive that, the kernel-based loss function $\mathcal{L}_{MEE}$ is bounded and approaches zero when the residual magnitude becomes large. As a result, the exponential of the negative loss function $\exp\left(-\mathcal{L}_{MEE}\right)$ exhibits a nonzero constant in the tail of the error space, and cannot be normalized into a proper distributional model using the normalization term that is equal to the integral $\int_e\exp\left(-\mathcal{L}_{MEE}\right)de$ over the entire support. Accordingly, MEE learning criterion is not interpreted here as defining a normalizable data distribution model to build a standard likelihood function for characterizing a probabilistic process. Instead, we formulate the proposed approach adopting the generalized Bayesian framework \cite{bissiri2016general}, also known as Gibbs posterior framework \cite{martin2022direct}, where the posterior updating process is driven by an exponentiated loss function, instead of standard likelihood term. To be specific, under the generalized Bayesian framework, the posterior update is defined via a loss-calibrated approach instead of using a standard likelihood-based posterior function. Given prior distributions $p(\textbf{w}|\textbf{a})$ and $p(\textbf{a})$ concerning the model parameter and relevance parameter, respectively, the generalized posterior for the MEE-based SBL is formulated as:
\begin{equation}
\label{Eq.PostSblMee}
\tilde{p}\left(\textbf{w},\textbf{a}|\textbf{t}\right)\propto\exp\left(-\mathcal{L}_{MEE}\right)p(\textbf{w}|\textbf{a})p(\textbf{a})
\end{equation}
where the tilde symbol indicates that $\tilde{p}\left(\textbf{w},\textbf{a}|\textbf{t}\right)$ is a generalized posterior controlled by the MEE-induced empirical risk $\mathcal{L}_{MEE}$ rather than a conventional likelihood. Thus, sparsity is encoded through the SBL prior structure, while robustness is introduced using the MEE criterion. Equivalently the generalized posterior update in (\ref{Eq.PostSblMee}) can be considered as balancing an empirical risk minimization and Kullback-Leibler divergence from the priors. Specifically, from a variational perspective, given an auxiliary distribution $q\left(\textbf{w},\textbf{a}\right)$, the generalized posterior in (\ref{Eq.PostSblMee}) suggests the following optimization task over this auxiliary distribution:
\begin{equation}
\begin{split}
\label{Eq.PostSblMee2}
q^*&\left(\textbf{w},\textbf{a}\right)=\arg\underset{q\left(\textbf{w},\textbf{a}\right)}{\min}\\
&\left\{\mathbb{E}_{q\left(\textbf{w},\textbf{a}\right)}\left[\mathcal{L}_{MEE}\right]+KL\left[q\left(\textbf{w},\textbf{a}\right)||p(\textbf{w}|\textbf{a})p(\textbf{a})\right]\right\}\\
\end{split}
\end{equation}
which highlights a trade-off between robust fitting and sparsity regularization \cite{bissiri2016general,martin2022direct}.

In practice, a direct inference under the generalized posterior defined in (\ref{Eq.PostSblMee}) is generally intractable. Hence we aim to obtain the solution that maximizes the generalized posterior form, i.e., a generalized MAP estimate. To this end, variational inference is adopted to approximate this intractable generalized posterior \cite{blei2017variational}. Specifically, we leverage an auxiliary distribution $q\left(\textbf{w},\textbf{a}\right)$ to approach the generalized posterior $\tilde{p}\left(\textbf{w},\textbf{a}|\textbf{t}\right)$ by maximizing the generalized evidence lower bound (ELBO) in the following expression:
\begin{equation}
\label{Eq.Elbo1}
\max ELBO(q\left(\textbf{w},\textbf{a}\right))=\mathbb{E}_{q\left(\textbf{w},\textbf{a}\right)}\left[\log\frac{\tilde{p}\left(\textbf{w},\textbf{a}|\textbf{t}\right)}{q\left(\textbf{w},\textbf{a}\right)}\right]
\end{equation}
To facilitate the variational inference, we further employ mean-field factorization to estimate $\textbf{w}$ and $\textbf{a}$, respectively, as follows:
\begin{equation}
\label{Eq.MeanField}
q\left(\textbf{w},\textbf{a}\right)=q_{\textbf{w}}(\textbf{w})q_{\textbf{a}}(\textbf{a})
\end{equation}
where $q_{\textbf{w}}(\textbf{w})$ and $q_{\textbf{a}}(\textbf{a})$ denote the surrogate models to estimate the MAP results for $\textbf{w}$ and $\textbf{a}$, respectively. Thus, the variational inference for the MEE-based SBL can be expressed as follows:
\begin{equation}
\label{Eq.FreeEnergy}
\max ELBO=\mathbb{E}_{q_{\textbf{w}}(\textbf{w})q_{\textbf{a}}(\textbf{a})}\left[\log\frac{\tilde{p}\left(\textbf{w},\textbf{a}|\textbf{t}\right)}{q_{\textbf{w}}(\textbf{w})q_{\textbf{a}}(\textbf{a})}\right]
\end{equation}
In variational inference method, one could obtain the following two equations which can alternately maximize the ELBO value
\begin{equation}
\label{Eq.FreeEnergyAlt}
\begin{split}
\log q_{\textbf{w}}(\textbf{w})&=\mathbb{E}_{q_{\textbf{a}}(\textbf{a})}\left[\log \tilde{p}\left(\textbf{w},\textbf{a}|\textbf{t}\right)\right]+const.\\
\log q_{\textbf{a}}(\textbf{a})&=\mathbb{E}_{q_{\textbf{w}}(\textbf{w})}\left[\log \tilde{p}\left(\textbf{w},\textbf{a}|\textbf{t}\right)\right]+const.\\
\end{split}
\end{equation}
Through substituting in the generalized posterior (\ref{Eq.PostSblMee}), one then obtains:
\begin{equation}
\label{Eq.LogQw}
\log q_{\textbf{w}}(\textbf{w})=\sum_{i=1}^{N}\sum_{j=1}^{M}\eta_j\cdot k_{\sigma}\left(e_i-c_j\right)-\frac{1}{2}\textbf{w}^\top\mathbb{E}_{q_{\textbf{a}}(\textbf{a})}\left[\textbf{A}\right]\textbf{w}
\end{equation}
\begin{equation}
\label{Eq.LogQa}
\log q_{\textbf{a}}(\textbf{a})=\sum_{d=1}^{D}\left(-\frac{1}{2}a_d\mathbb{E}_{q_{\textbf{w}}(\textbf{w})}\left[w_d^2\right]-\frac{1}{2}\log a_d\right)
\end{equation}
where $\textbf{A}=diag\left(a_1,\cdots,a_D\right)\in\mathbb{R}^{D\times D}$ indicates the diagonal precision matrix, and the constants are discarded for simplicity.

First, one could optimize the distribution $q_{\textbf{w}}(\textbf{w})$ with a fixed distribution $q_{\textbf{a}}(\textbf{a})$, that the mathematical expectation $\mathbb{E}_{q_{\textbf{a}}(\textbf{a})}\left[\textbf{A}\right]$ is known. However, because (\ref{Eq.LogQw}) is not a quadratic expression, $q_{\textbf{w}}(\textbf{w})$ cannot be analytically formed as a Gaussian distribution as conventional variational inferences. To address this obstacle, we further leveraged the Laplacian approximation method that approximates $\log q_{\textbf{w}}(\textbf{w})$ by the following quadratic expression:
\begin{equation}
\label{Eq.LogQwLap}
\log q_{\textbf{w}}(\textbf{w})\approx\log q_{\textbf{w}}(\textbf{w}^*)-\frac{\left(\textbf{w}-\textbf{w}^*\right)^\top \textbf{H}\left(\textbf{w}^*\right)\left(\textbf{w}-\textbf{w}^*\right)}{2}
\end{equation}
in which $\textbf{w}^*$ is the maximum point of $\log q_{\textbf{w}}(\textbf{w})$, while $\textbf{H}\left(\textbf{w}^*\right)$ represents the negative Hessian matrix for $\log q_{\textbf{w}}(\textbf{w})$ evaluated at $\textbf{w}^*$. Thus, $q_{\textbf{w}}(\textbf{w})$ is approximated by a Gaussian distribution:
\begin{equation}
\label{Eq.LogQwLap2}
q_{\textbf{w}}(\textbf{w})\approx\mathcal{N}\left(\textbf{w}|\textbf{w}^*,\textbf{H}\left(\textbf{w}^*\right)^{-1}\right)
\end{equation}
To acquire the optimal parameter $\textbf{w}^*$ that maximizes $\log q_{\textbf{w}}(\textbf{w})$ for Laplacian approximation, one may notice that the objective function in (\ref{Eq.LogQw}) is equal to $L_2$-regularized MEE objective, with $\textbf{A}$ denoting the penalty coefficient, where one can use gradient-based optimization methods. In particular, for linear regression one can utilize the fixed-point approach since the model output $\hat{t}$ is linear with respect to model parameter $\textbf{w}$ \cite{zhang2015convergence}. On the other hand, for logistic regression, half-quadratic technique provides an effective way for optimizing the model parameter regarding MEE-based classification \cite{li2021restricted}. The optimization procedure for obtaining $\textbf{w}^*$ which maximizes $\log q_{\textbf{w}}(\textbf{w})$ in (\ref{Eq.LogQw}) is elaborated in Appendix C (see supplementary material). After calculating the optimal model parameter $\textbf{w}^*$, we could acquire the negative Hessian matrix in (\ref{Eq.Hessian}), in which $\frac{\partial\hat{t_i}}{\partial\textbf{w}}$ and $\frac{\partial^2\hat{t_i}}{\partial\textbf{w}\partial\textbf{w}^\top}$ are dependent on the specific configuration of the utilized link function in (\ref{Eq.Glm1}).

\begin{figure*}[hbt]
\centering
\begin{equation}
\label{Eq.Hessian}
\textbf{H}\left(\textbf{w}\right)=-\sum_{i=1}^{N}\sum_{j=1}^{M}\frac{\eta_j}{\sigma^2}\exp\left(-\frac{\left(e_i-c_j\right)^2}{2\sigma^2}\right)\left[\left(\frac{\left(e_i-c_j\right)^2}{\sigma^2}-1\right)\frac{\partial\hat{t_i}}{\partial\textbf{w}}\left(\frac{\partial\hat{t_i}}{\partial\textbf{w}}\right)^\top+\left(e_i-c_j\right)\frac{\partial^2\hat{t_i}}{\partial\textbf{w}\partial\textbf{w}^\top}\right]+\mathbb{E}_{q_{\textbf{a}}(\textbf{a})}\left[\textbf{A}\right]
\end{equation}
\hrulefill
\end{figure*}

Thus, after optimizing the distribution $q_{\textbf{w}}(\textbf{w})$, we then focus on optimizing the distribution $q_{\textbf{a}}(\textbf{a})$ in (\ref{Eq.LogQa}). Notably, one could perceive that, $q_{\textbf{a}}(\textbf{a})$ exhibits the following Gamma distribution by performing an exponential function on $\log q_{\textbf{a}}(\textbf{a})$ in (\ref{Eq.LogQa}) as:
\begin{equation}
\label{Eq.Qa}
\begin{split}
q_{\textbf{a}}(\textbf{a})&=\prod_{d=1}^{D}\exp\left(-\frac{1}{2}\mathbb{E}_{q_{\textbf{w}}(\textbf{w})}\left[w_d^2\right]a_d-\frac{1}{2}\log a_d\right)\\
&\propto\prod_{d=1}^{D}\varGamma\left(a_d|\frac{1}{2},\frac{1}{2}\mathbb{E}_{q_{\textbf{w}}(\textbf{w})}\left[w_d^2\right]\right)\\
\end{split}
\end{equation}
where $\varGamma\left(a_d|\frac{1}{2},\frac{1}{2}\mathbb{E}_{q_{\textbf{w}}(\textbf{w})}\left[w_d^2\right]\right)$ indicates a Gamma distribution with respect to $a_d$ parameterized by the shape parameter $\frac{1}{2}$ and the rate parameter $\frac{1}{2}\mathbb{E}_{q_{\textbf{w}}(\textbf{w})}\left[w_d^2\right]$. Since we have approximated $q_{\textbf{w}}(\textbf{w})$ through a Gaussian distribution in (\ref{Eq.LogQwLap2}), the expectation $\mathbb{E}_{q_{\textbf{w}}(\textbf{w})}\left[w_d^2\right]$ can be easily calculated by the following equation:
\begin{equation}
\label{Eq.ExpecW}
\mathbb{E}_{q_{\textbf{w}}(\textbf{w})}\left[w_d^2\right]=w^{*2}_d+[\textbf{H}\left(\textbf{w}^*\right)^{-1}]_{d,d}
\end{equation}
in which the second term of right-hand side represents the $d$-th diagonal element of $\textbf{H}\left(\textbf{w}^*\right)^{-1}$. Consequently, the optimization for $q_{\textbf{a}}(\textbf{a})$ is achieved, and the expectation $\mathbb{E}_{q_{\textbf{a}}(\textbf{a})}\left[\textbf{A}\right]$ in (\ref{Eq.LogQw}) can be naturally calculated with the updated $q_{\textbf{a}}(\textbf{a})$ by the following equation:
\begin{equation}
\label{Eq.UpdateASlow}
\mathbb{E}_{q_{\textbf{a}}(\textbf{a})}\left[a_d\right]=\frac{1}{\mathbb{E}_{q_{\textbf{w}}(\textbf{w})}\left[w_d^2\right]}=\frac{1}{w^{*2}_d+[\textbf{H}\left(\textbf{w}^*\right)^{-1}]_{d,d}}
\end{equation}
To accelerate the parameter convergence, one can alternatively utilize the following rule for updating the relevance parameters
\begin{equation}
\label{Eq.UpdateAFast}
\mathbb{E}_{q_{\textbf{a}}(\textbf{a})}\left[a_d\right]=\frac{1-\mathbb{E}_{q_{\textbf{a}}(\textbf{a})}\left[a_d\right]\cdot[\textbf{H}\left(\textbf{w}^*\right)^{-1}]_{d,d}}{w^{*2}_d}
\end{equation}
which was derived by the effective number of parameters \cite{mackay1992bayesian}. The updated value of $\mathbb{E}_{q_{\textbf{a}}(\textbf{a})}\left[a_d\right]$ is further substituted into (\ref{Eq.LogQw}), so as to optimize $\log q_{\textbf{w}}(\textbf{w})$ again, thus effectuating an iterative optimization procedure to maximize ELBO through variational inference.

\begin{algorithm}[h]
\caption{\textit{SBL-MEE (see detailed version in Appendix D)}}
\label{SBL-MEE}
\begin{algorithmic}[1]		
\State \textbf{input}:

Training samples $\left\{\left(\textbf{x}_i,t_i\right)\right\}_{i=1}^N$;

Kernel bandwidth $\sigma$;

Pruning threshold $a_{max}$;
		
\State \textbf{parameter setting}:

For regression, once the model parameter is changed, update the quantization element $c_j$ and weight $\eta_j$ utilizing Algorithm \ref{Algo.Quantization};

For classification, first employ a preliminary classifier to obtain the prediction errors $e_i=t_i-\hat{t_i}$. Then, determine the quantization weight $\eta$ using (\ref{Eq.RmeeHyper});

\Repeat 
\State $\textbf{w}$\textit{-step}: update $\textbf{w}$ according to Appendix C;
\State $\textbf{a}$\textit{-step}: update $\textbf{a}$ according to (\ref{Eq.UpdateAFast});
\If{$a_d\geqslant a_{max}$}
\State prune the corresponding dimension from the model training process and also set the model parameter $w_d=0$;
\EndIf
\Until the increase for ELBO value (\ref{Eq.FreeEnergy}) is sufficiently small or the number of iterations exceeds upper constraint value;

\State \textbf{MAP estimation}:

Acquire the optimal model parameter that maximizes $\log q_{\textbf{w}}(\textbf{w})$ in (\ref{Eq.LogQw}) utilizing the fixed expectation $\mathbb{E}_{q_{\textbf{a}}(\textbf{a})}\left[\textbf{A}\right]$;

\State \textbf{output}:
		
Model parameter $\textbf{w}\in\mathbb{R}^D$.

\end{algorithmic}
\end{algorithm}

After accomplishing the convergence of parameter learning, the surrogate models $q_{\textbf{w}}(\textbf{w})$ and $q_{\textbf{a}}(\textbf{a})$ can approximate the true posterior distributions regarding $\textbf{w}$ and $\textbf{a}$, respectively, through maximizing the ELBO (\ref{Eq.FreeEnergy}). Then, to obtain an adequate model parameter for regression or classification, one can simply adopt an MAP estimation from the surrogate model $q_{\textbf{w}}(\textbf{w})$ associated with the fixed $\mathbb{E}_{q_{\textbf{a}}(\textbf{a})}\left[\textbf{A}\right]$. This robust SBL approach using MEE proposed in this study is named as SBL-MEE, which is briefly summarized in Algorithm \ref{SBL-MEE}. Detailed implementation of SBL-MEE is specified in Appendix D (see supplementary material).

\section{Experiments}
\label{sec:results}

To improve the brain activity decoding performance in noisy and small-dataset scenarios, this paper developed a reorganized SBL framework using the robust MEE learning criterion rather than conventional likelihood model. To systematically evaluate our developed SBL-MEE framework, this paper employed two real-world brain decoding tasks of regression and classification contexts, respectively. In performance comparison, we adopted the conventional SBL implementations that employ a Gaussian likelihood for regression and a Bernoulli likelihood for logistic regression. Moreover we also leveraged the recently developed SBCL framework \cite{li2023adaptive,li2023correntropy,li2025sparse,li2025correntropy} in performance comparison which represents the state-of-the-art technique for robust sparse brain decoding. Furthermore, we compared our proposed SBL-MEE with two traditional while effective methods, including the $L_2$-regularized ridge algorithm (using an isotropic Gaussian prior) and the $L_1$-regularized lasso algorithm (using a Laplace prior). Through the following experiment, the pruning threshold $a_{max}$ was fixed as $10^6$ for each ARD-based sparse method, including the traditional SBL, SBCL, and SBL-MEE. The regularization coefficients for ridge and lasso algorithms were determined via cross validation. The maximal iteration was set to 500 for each method. Other hyperparameters are described for each dataset, respectively.

\begin{figure*}[htb]
\centering
\includegraphics[width=0.912\textwidth]{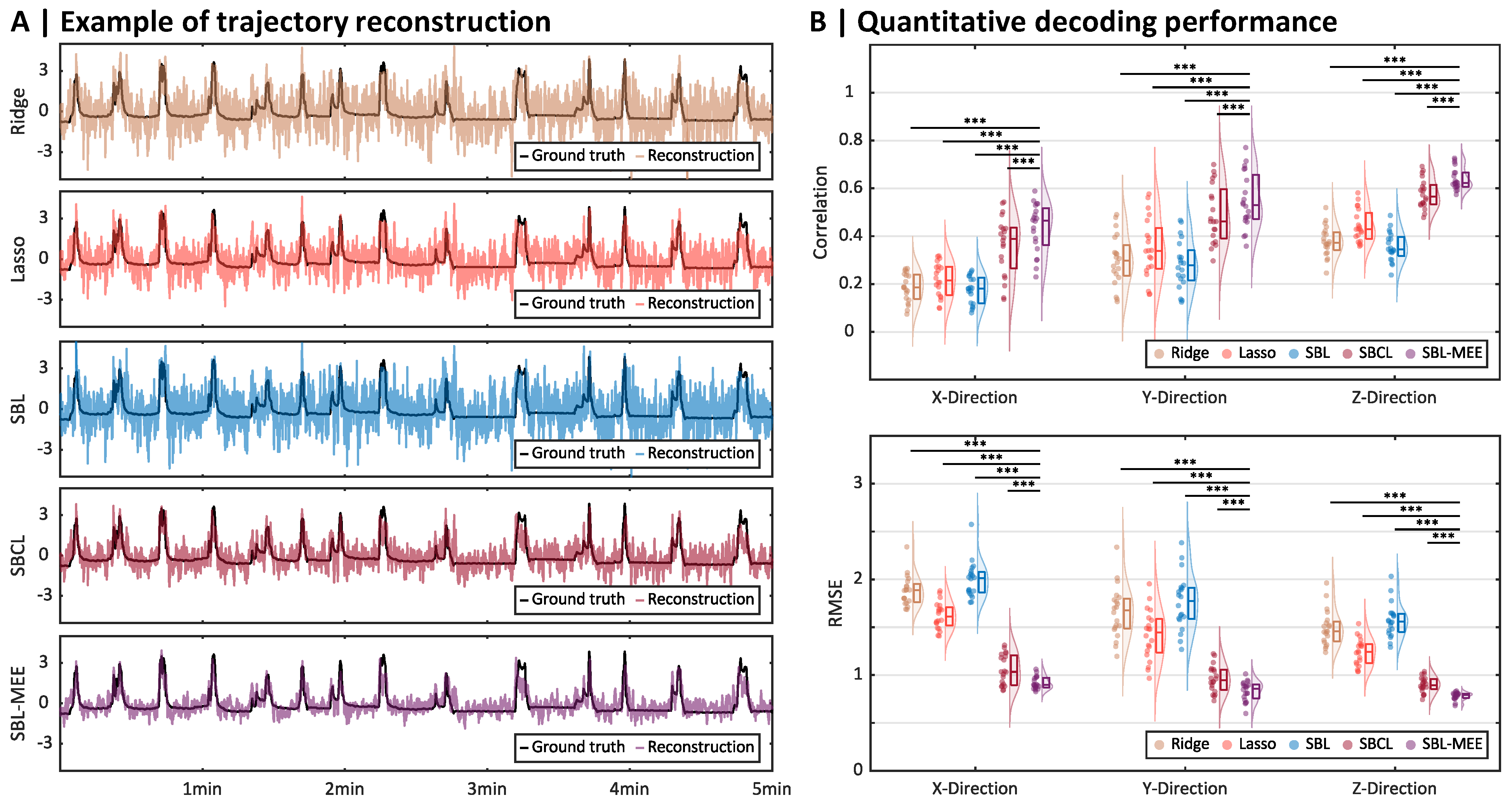}
\vspace{-2mm}
\caption{ECoG-based movement trajectory reconstruction task: (A) example of the comparison between original and reconstructed movement trajectory decoded by different approaches (Y-Direction in Session No.9 for Monkey B); (B) the quantitative decoding performance of different approaches with three movement directions, examined by a non-parametric Friedman test and post-hoc comparison with the Bonferroni corrections ($n=20$ sessions for each direction in movement trajectory decoding, ${}^{***}p<0.001$).}
\label{FigResEcog}
\end{figure*}

\begin{figure}[htb]
\centering
\includegraphics[width=0.912\columnwidth]{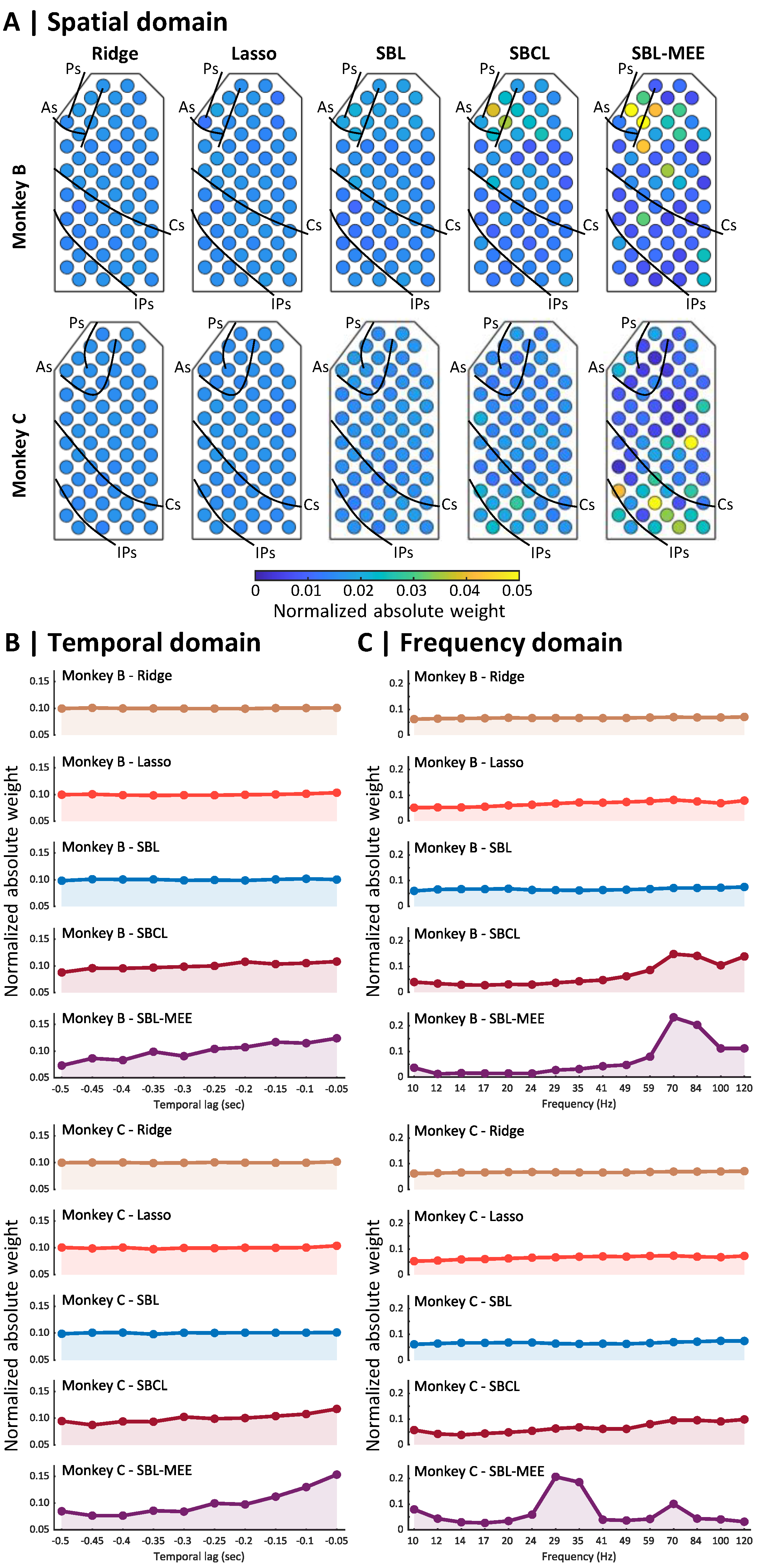}
\caption{Physiological patterns for trajectory reconstruction ave-raged across ten sessions and three directions for each monkey.}
\label{FigResEcogFeature}
\end{figure}

\subsection{Regression Task: ECoG-Based Movement Trajectory Reconstruction}

This study first utilized a real-world brain decoding scenario for performance evaluation on regression which aims to realize reconstruction of continuous movement trajectory using ECoG signal \cite{shimoda2012decoding}. In this dataset, two macaques were trained to track foods using right hands, with the continuous three-dimensional trajectory of the right hand recorded. Each macaque performed ten sessions of 15 minutes, with the brain activity concurrently measured via a 64-ch ECoG array. The trajectories were down-sampled to 10 Hz, thus yielding 9,000 samples in each session. We employed the same ECoG-based decoding protocol as \cite{shimoda2012decoding} where the trajectory was reconstructed via a 9,600-dimensional wavelet time-frequency feature. Other details for the employed dataset, preprocessing steps, feature construction and decoding framework are elaborated in Appendix E.1. (see supplementary material)

Concerning the hyperparameter settings, for both SBCL and SBL-MEE, the kernel bandwidth was determined by a five-fold cross validation in the training set, in which the optimal kernel bandwidth exhibited the highest average correlation coefficient in cross validation. For the quantization process of SBL-MEE, the threshold $\varepsilon$ was set as $\frac{\max(e)-\min(e)}{20}$, in which $e$ represents the current residual, thus leading to no more than 20 codebook elements.

First, we followed the default decoding setting as \cite{shimoda2012decoding} where the model was trained with the first ten minutes of each session and then evaluated via the last five minutes in the same session. Hence 6,000 training samples (10 min $\times$ 10 Hz) were available to handle the 9,600-dimensional feature. Based on this setting, Fig. \ref{FigResEcog}(A) illustrates an example for visual comparison between the ground-truth and reconstructed hand movement trajectories decoded via different approaches. From the visual observation, one can perceive that, robust SBL approaches including SBCL and SBL-MEE realized evidently more accurate reconstruction than those traditional methods. Remarkably the proposed SBL-MEE demonstrated superior fidelity in reconstructing the hand movement trajectories than SBCL, as evidenced by the smaller discrepancies from the ground truth. To quantitatively compare the decoding performance, we further computed the correlation coefficients and root mean squared errors (RMSE) between the original and reconstructed trajectories. Fig. \ref{FigResEcog}(B) illustrates the quantitative decoding performance of each method on different movement directions. One can observe that, regarding all three directions, the proposed SBL-MEE method showed the highest correlation and the lowest RMSE value. Moreover, our method outperformed the other four existing methods with statistically significant differences according to a non-parametric Friedman test.

Next, we investigated the physiological pattern in movement reconstruction revealed by each approach, which is particularly important for interpretation and evaluation with neuroscientific knowledge. Therefore we further illustrate the distribution map of decoder weight obtained from each method, evaluating their plausibility on movement decoding. To be specific, considering the decoder weight $w_{ch,temp,freq}$ that corresponds to electrode $ch$, temporal lag $temp$, and frequency band $freq$, we evaluated the normalized absolute weight (NAW) of each feature through
\begin{equation}
\label{Eq.ModelContrib1}
Naw(ch)=\frac{\sum_{temp}\sum_{freq}|w_{ch,temp,freq}|}{\sum_{ch}\sum_{temp}\sum_{freq}|w_{ch,temp,freq}|}
\end{equation}
\begin{equation}
\label{Eq.ModelContrib2}
Naw(temp)=\frac{\sum_{ch}\sum_{freq}|w_{ch,temp,freq}|}{\sum_{ch}\sum_{temp}\sum_{freq}|w_{ch,temp,freq}|}
\end{equation}
\begin{equation}
\label{Eq.ModelContrib3}
Naw(freq)=\frac{\sum_{ch}\sum_{temp}|w_{ch,temp,freq}|}{\sum_{ch}\sum_{temp}\sum_{freq}|w_{ch,temp,freq}|}
\end{equation}
where each numerator sums the absolute decoder weight which corresponds to one specific feature dimension in every domain, while the denominator calculates the sum of the absolute value of all decoder weights, serving as a normalization factor. Thus, NAW is appropriate for characterizing the relative distributions of decoder emphasis across each feature domain. Based on this measure, we visualize the decoder weight map of each method in Fig. \ref{FigResEcogFeature} for each domain, and evaluate which algorithm shows a better weight map that is more consistent with neuroscientific knowledge, providing a basis for evaluating the interpretability. Note that a physiologically plausible pattern should not merely be sparse but also highlight task-relevant channels, meaningful temporal lags, and frequency bands known to encode the motor information in this movement decoding task. First, considering the spatial patterns presented in Fig. \ref{FigResEcogFeature}(A), for Monkey B, SBL-MEE exhibited a clear concentration in the dorsal-anterior part of the ECoG array, in particular around the Arcuate sulcus (As) and the Principal sulcus (Ps), with additional distributions near the precentral region adjacent to the Central sulcus (Cs). These areas are closely related to motor processing in the non-human primate, where the dorsal premotor cortex around As is critical for movement planning and preparation, the regions around Ps are relevant to prefrontal cortical processing, and the precentral area around Cs is involved in movement execution. In contrast, Ridge, Lasso, and SBL showed relatively diffuse spatial maps, with weights distributed broadly across the electrode array and without a clear anatomical focus. Although SBCL also showed a localized focus, the spatial organization remained less legible than that of SBL-MEE. On the other hand, concerning Monkey C, SBL-MEE assigned larger weights around Cs and extending toward the intraparietal sulcus (IPs). Although this pattern map is distinct from that of Monkey B, it is also meaningful because the cortical regions near Cs are highly relevant to sensorimotor and movement-execution process, while the regions around IPs contribute to visuomotor transformation and movement-related spatial representations in non-human primates. By comparison, other methods consistently produced an overly smooth decoder pattern on Monkey C with limited anatomical specificity. Next, regarding the temporal patterns illustrated in Fig. \ref{FigResEcogFeature}(B), one can observe that each approach produced similar maps across both macaques. Specifically, Ridge, Lasso, and SBL exhibited a flat distribution across different lags, indicating a limited temporal emphasis in decoder weight. SBCL revealed a modest increase on lags closer to movement onset, suggesting partial sensitivity to the motor task. In contrast, SBL-MEE consistently exhibited a more structured temporal map, with a relatively larger weight assigned to the more recent lag. Such a distribution is plausible physiologically, because neural activity closer to the movement is more directly related to motor preparations and the transition to execution. Hence, rather than assigning a similar importance to each lag, SBL-MEE emphasized a limited temporal window, which is more likely to contain the motor-relevant information. Finally, considering the frequency patterns in Fig. \ref{FigResEcogFeature}(C), Ridge, Lasso, and SBL generally exhibited flat frequency distribution, while SBCL revealed only modest spectral preferences on both macaques. By comparison, SBL-MEE showed structured maps in frequency domain. For Monkey B, the dominant weight was concentrated in the high-gamma band, which is consistent with the well-established role of high-frequency ECoG components in encoding movement-related cortical information. Regarding Monkey C, SBL-MEE exhibited a dominant focus around low-gamma and high-beta bands with a secondary peak in the high-gamma range. This pattern indicates that, SBL-MEE preserved both the intermediate-frequency sensorimotor components and the high-frequency motor-related information, instead of using similar weight across the entire frequency range. Overall, these patterns suggest that SBL-MEE does not merely enforce sparse solutions but also recognize a more physiologically meaningful decoder, thereby providing improved interpretability for neural decoding.

\begin{figure}[t]
\centering
\includegraphics[width=0.98\columnwidth]{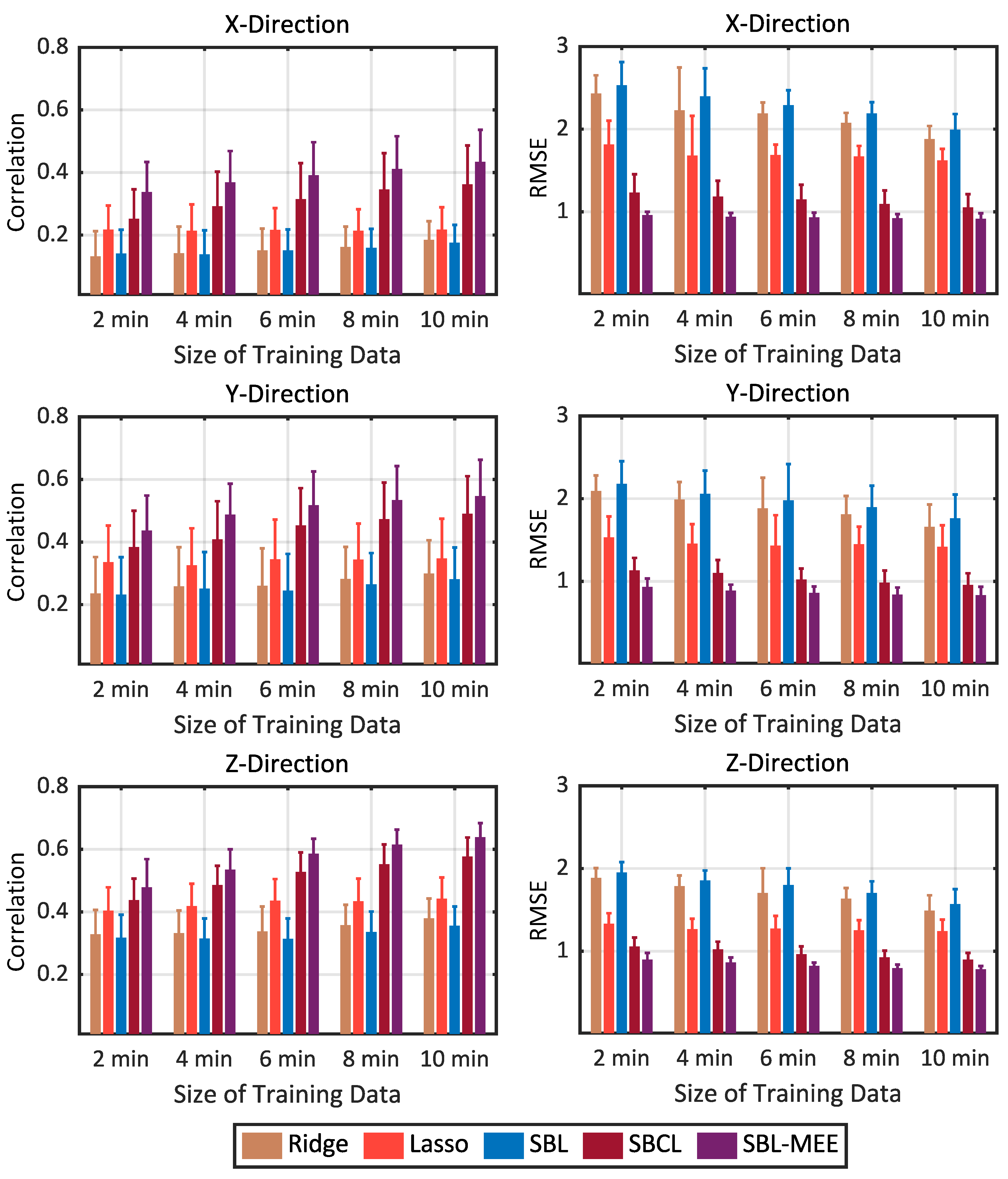}
\vspace{-2mm}
\caption{Movement reconstruction performance for each method based on different sizes of training data averaged across twenty sessions. Error bars indicate the associated standard deviations.}
\label{FigResEcogFew}
\end{figure}

Furthermore, to evaluate each decoding method with a more challenging condition, we further decreased the size of training data in movement reconstruction. Specifically, for each session that lasted 15 minutes, we also leveraged the first 2 min, 4 min, 6 min, and 8 min, respectively, to train each method and tested them on the remaining part of each session. For example, when we employed 2 min in training, each model only utilized 1,200 training samples to handle 9,600 dimensions, yielding a highly challenging high-dimensional low-sample-size problem. Fig. \ref{FigResEcogFew} presents the movement reconstruction performance of different methods in this more challenging setting. One can observe that even though using a smaller number of training samples, SBL-MEE consistently outperformed other algorithms in movement reconstruction with higher correlation and lower RMSE values. This result further demonstrates the superiority of the proposed SBL-MEE for handling high-dimensional and low-sample-size problems.

\subsection{Classification Task: fMRI-Based Visual Stimulus Reconstruction}

Furthermore, concerning the classification context, this work evaluated different decoding methods, adopting an fMRI-based visual stimulus reconstruction task \cite{miyawaki2008visual}. During the experiment one human subject viewed visual stimulus composed of 10$\times$10 contrast-based patch images, yielding 100 binary classification tasks for pixel-wise reconstructions. This experiment consisted of two sessions, including the random image session for model training with 440 random images, and the figure image session for performance evaluation that utilized three image categories of 40 blocks within each category. In reconstruction, 100 pixel-wise classifiers were first trained via different algorithms using the block-averaged fMRI of the visual cortex, and then linearly combined to recover the stimuli. A detailed description for this task is provided in Appendix E.2. (see supplementary material)

For the hyperparameter settings, the weighting coefficient $\eta$ for SBL-MEE, as denoted in (\ref{Eq.RmeeHyper}), was determined by utilizing SBCL as a preliminary model, i.e. the prediction error $e$ in (\ref{Eq.RmeeHyper}) was obtained from SBCL. Regarding the kernel bandwidth for SBCL and SBL-MEE, since the visual stimulus reconstruction task adopted a relatively complicated training process, it would be difficult to select the individually optimal kernel bandwidth for each pixel-wise classifier using cross validation. Hence, we applied a uniform value to the kernel bandwidth for both SBL-MEE and SBCL. As suggested in \cite{li2023correntropy,li2021restricted}, one could set the kernel bandwidth to be 1.0 that realized a satisfactory trade-off between model robustness and training stability in both SBCL and RMEE based classifications. Therefore, we used this value for SBL-MEE and SBCL algorithms through the classification context.

\begin{figure*}[t]
\centering
\includegraphics[width=0.912\textwidth]{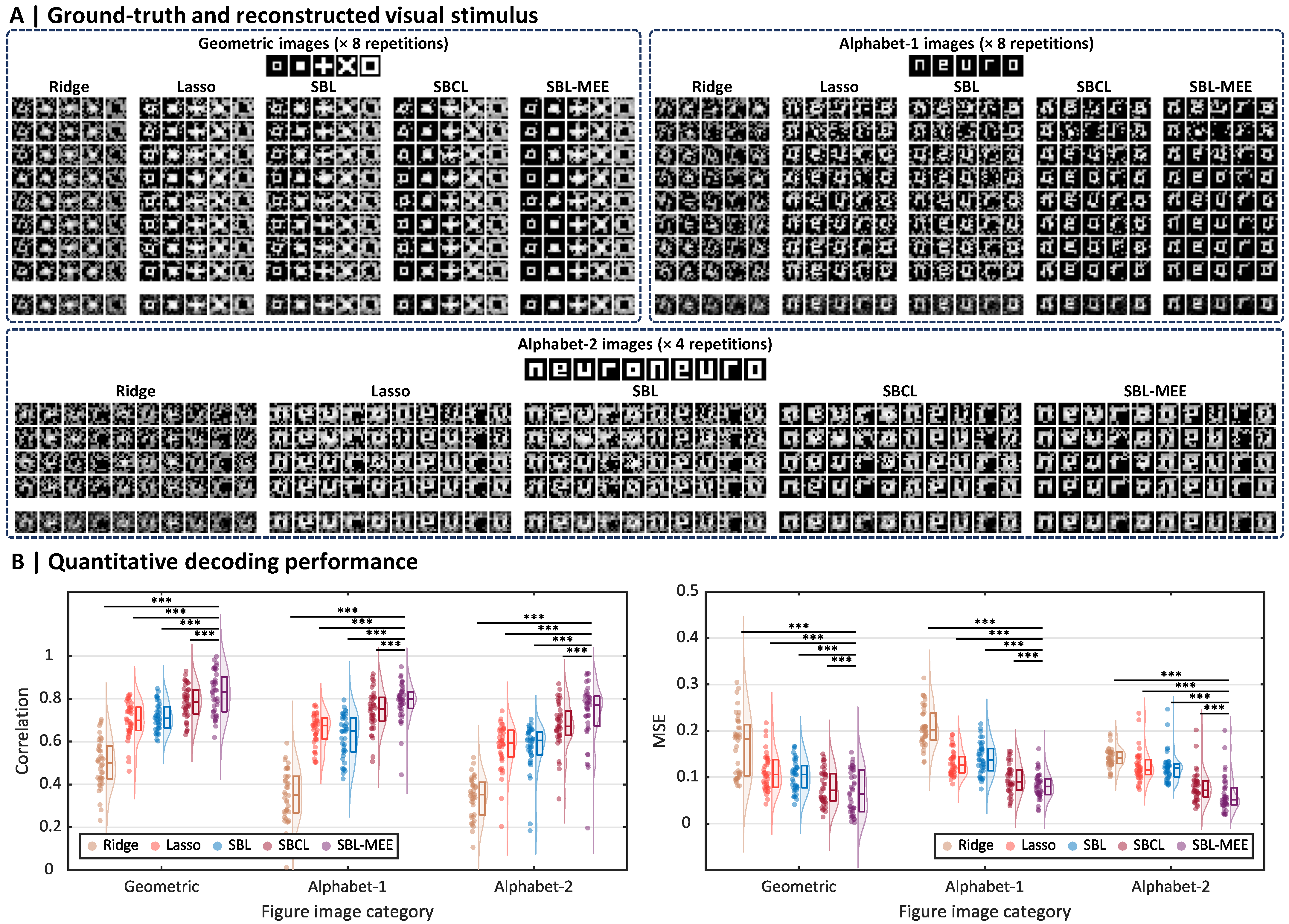}
\vspace{-2mm}
\caption{fMRI-based visual stimulus reconstruction task: (A) a comparison between the original and reconstructed visual stimulus by different approaches, where the bottom rows illustrate the reconstructions averaged across repetitions for each image category; (B) quantitative decoding performance for each approach, examined by a non-parametric Friedman test and post-hoc comparison associated with Bonferroni corrections ($n=40$ stimulus blocks in each of three different figure image categories, ${}^{***}p<0.001$).}
\label{FigResVisual}
\end{figure*}

First, we adopted the original decoding setting as \cite{miyawaki2008visual} where the visual stimulus was reconstructed from the fMRI recording in V1 and V2 regions with 1,698 voxels. As a result, classifiers were trained using 440 training images with 1,698-dimensional features. Under this protocol, the reconstructed visual stimulus considering each block in the figure image session is illustrated in Fig. \ref{FigResVisual}(A) compared to the ground truth. On visual inspection one can perceive that, the reconstructed stimulus by SBL-MEE exhibited a more legible pattern similar to the original stimulus compared to that decoded by other approaches. In addition, the decoding performance for each method was also quantitatively evaluated through measuring the correlation and MSE between the original and reconstructed visual stimulus. Fig. \ref{FigResVisual}(B) shows the quantitative decoding performance for different approaches obtained on the 40 blocks regarding different image categories. To investigate the statistical difference between each algorithm we also utilized the non-parametric Friedman test and post-hoc pairwise comparisons with the Bonferroni corrections. One can perceive that, for three different image categories, the proposed SBL-MEE realized the highest correlation and the lowest MSE for the visual reconstruction. Further, SBL-MEE outperformed the other four methods with statistically significant differences, associated with considerably small $p$-values. This classification scenario demonstrates moreover the advantage of our proposed approach in real-world high-dimensional brain signal decoding applications.

\begin{figure*}[t]
\centering
\includegraphics[width=1\textwidth]{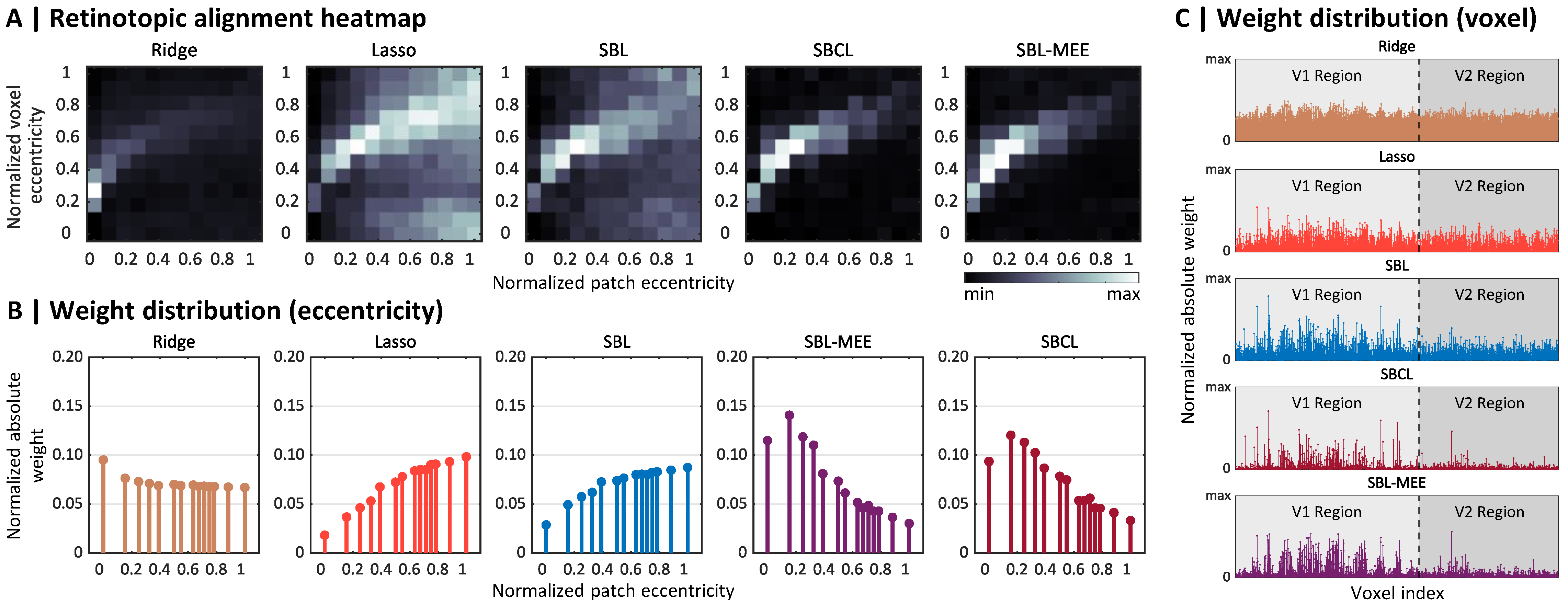}
\caption{Physiological patterns on visual stimulus reconstruction averaged across 100 pixel-wise classifiers obtained from different algorithms: (A) retinotopic alignment heatmap shows the normalized absolute weight with respect to each pairwise combination for voxel eccentricity and patch eccentricity; (B) patch-eccentricity-wise weight distribution; (C) voxel-wise weight distribution.}
\label{FigResVisualFeature}
\end{figure*}

\begin{figure*}[htb]
\centering
\includegraphics[width=0.912\textwidth]{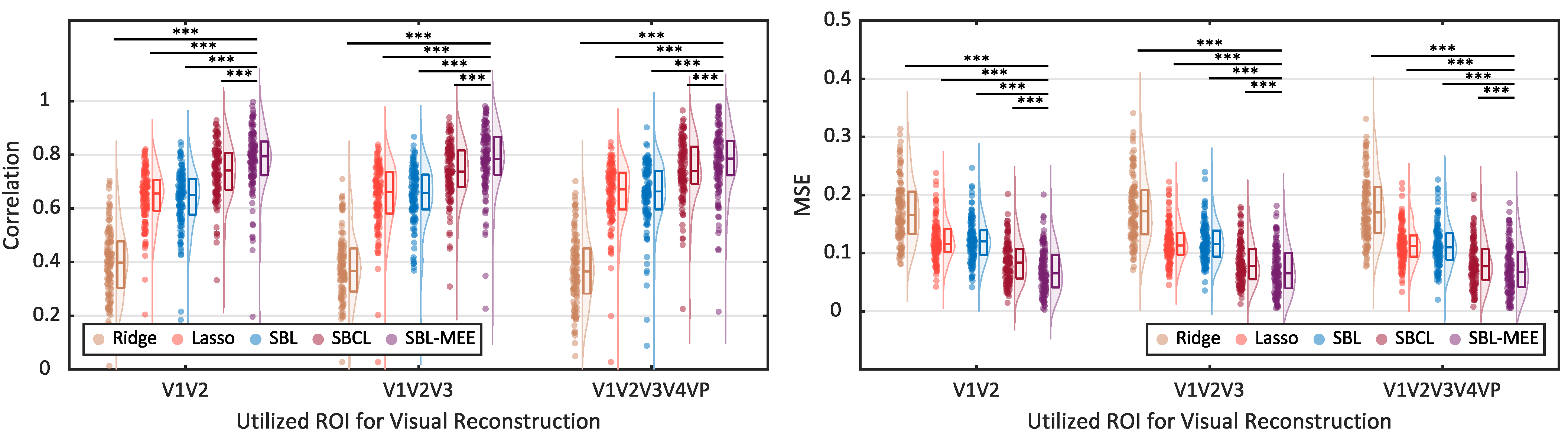}
\vspace{-2mm}
\caption{Visual reconstruction performance of different methods on larger regions of interest (ROI), examined by a non-parametric Friedman test and post-hoc comparison with the Bonferroni corrections ($n=120$ blocks for all image categories, ${}^{***}p<0.001$).}
\label{FigResVisualFew}
\end{figure*}

Similarly we also studied the physiological patterns revealed by decoding weight for interpretation and evaluation regarding neuroscientific knowledge in this visual reconstruction context. Notably, visual information relevant to stimulus reconstruction is generally distributed across populations of voxels rather than being encoded in merely few units. Therefore a physiologically plausible pattern should not only maximize sparsity, but should reveal structured weight distribution aligned with known visual cortical organization. First, we present the normalized absolute weight for each pairwise combination of patch eccentricity and voxel eccentricity to visualize the retinotopic alignment in Fig. \ref{FigResVisualFeature}(A), which is a standard neuroscientific evaluation to examine whether the decoder weight follows the eccentricity-dependent retinotopic mapping known in the visual cortex \cite{miyawaki2008visual}. As shown in Fig. \ref{FigResVisualFeature}(A), Ridge revealed relatively weak diagonal structure, whereas Lasso and SBL presented broader while overly spread alignment map. By contrast, SBCL and particularly SBL-MEE demonstrated a more legible and compact diagonal distribution with high weight predominantly confined near the eccentricity-matched region rather than spreading across mismatched pairs. From a neuroscientific viewpoint, this pattern is desirable since the eccentricity matching is well-established for the retinotopic organization in visual cortex \cite{miyawaki2008visual}. Subsequently, Fig. \ref{FigResVisualFeature}(B) also illustrates the distribution for normalized absolute weights with respect to different patch eccentricities. Ridge revealed a rather flat trend across eccentricities, whereas Lasso and SBL showed increasing importance to more peripheral patches. On the other hand, SBCL and SBL-MEE exhibited a decreasing trend, using larger weights on low-eccentricity patch and gradually reduced contribution toward the periphery. This pattern is more rational physiologically, because central visual field representations are known to be over-represented in early visual cortex, associated with cortical magnification. Hence, the stronger weight of low-eccentricity patch suggests that SBCL and SBL-MEE are more consistent with the crucial effect of foveal and parafoveal areas for visual reconstruction. Finally, we also show the distribution of normalized absolute weight across each voxel in V1 and V2, respectively, in Fig. \ref{FigResVisualFeature}(C). One can observe that, Ridge revealed a relatively uniform distribution across each voxel for both two regions, while Lasso and SBL showed more fluctuations across voxels. However, no evident differentiation could be perceived between V1 and V2 for these two approaches. By comparison, SBCL and SBL-MEE produced more prominent concentration with stronger weights mainly distributed in V1 rather than V2. According to neuroscientific knowledge, this pattern has better physiological interpretability because the primary visual region V1 preserves more fine-grained spatial information relevant for visual reconstruction than V2 area. Therefore the V1-dominant allocation of SBCL and SBL-MEE suggests that these methods relied more on voxel populations consistent with the functional organization of the early visual cortex. In summary, the pattern learned by our SBL-MEE method exhibits strong physiological interpretability, which is also consistent with its desirable brain decoding performance, although SBCL also exhibited a similar pattern.

Furthermore, we also evaluated each approach based on this visual reconstruction dataset using a more challenging problem setting. To be specific, in addition to the default protocol which used V1 and V2 regions with 1,698 voxels, we further adopted two expanded regions of interest (ROI) in visual reconstruction to generate a more high-dimensional scenario, where V1V2V3 (2,721 voxels) and V1V2V3V4VP (3,412 voxels) were applied to this task, respectively. Fig. \ref{FigResVisualFew} shows the visual reconstruction performance of each method under this more high-dimensional problem setting. One can observe that our proposed SBL-MEE algorithm consistently realized the highest correlations and the lowest MSE values among the evaluated algorithms, associated with statistically significant difference compared with the other approaches, according to a non-parametric Friedman statistical test. This result indicates that our proposed method can achieve more effective brain activity decoding for the high-dimensional scenario.

\section{Discussion}
\label{sec:discussion}

In this study, we proposed a new robust SBL approach using the MEE learning criterion to construct a generalized posterior. The proposed SBL-MEE algorithm was evaluated by two brain activity decoding tasks including ECoG-based motor trajectory reconstruction (by regression), and fMRI-based visual stimulus reconstruction (by classification). Both two decoding scenarios consistently indicated that, our proposed SBL-MEE can realize superior brain decoding performance than the conventional and state-of-the-art algorithms, confirmed by the higher correlation and lower MSE metrics in the reconstruction for the behavioral and perceptual states. Hence, our approach provides a powerful tool for the development of BCI systems and the investigations of cognitive neuroscience, particularly for those problems with a restricted training dataset. In addition, SBL-MEE can capture a more physiologically plausible weight in brain decoding task, thereby enhancing the neuroscientific interpretability regarding predictions.

From a methodological viewpoint, this paper primarily aims to develop a robust and sparse GLM algorithm to address noisy and high-dimensional low-sample-size brain activity decoding. As reviewed in Appendix A, previous robust sparse approaches mainly utilize a sparsity-inducing regularization term on robust loss functions, such as $L_1$-regularized MCC or MEE. Although effective in practice, one has to determine two hyperparameters concurrently that control the robust loss and the model sparsity, respectively. This typically needs a cross validation over a two-dimensional grid of candidate hyperparameters, leading to high computational cost. Motivated by the ARD mechanism in SBL framework, our previous works, named SBCL \cite{li2023adaptive,li2023correntropy,li2025sparse,li2025correntropy}, made a preliminary attempt which incorporated the robust MCC into the SBL framework. These previous studies indicated that, this strategy not only achieves desirable model capability, but more importantly, avoids tuning the sparsity hyperparameter through the ARD mechanism. This significantly simplifies the practical implementation, as cross validation is performed over only one hyperparameter in loss function. This success further motivates our confidence in robust SBL framework. Since MEE has been proved with better applicability than MCC across different data distributions in ITL literature, it is natural to further extend our study by employing MEE, resulting in the proposed SBL-MEE algorithm. As expected this method realized superior decoding performance on brain datasets compared with SBCL. Although the principal difference between SBL-MEE and our previously developed SBCL lies in replacing MCC with MEE to construct the generalized posterior, this paper shows a deeper conceptual insight. In the SBCL, we still attempted to associate MCC with an underlying probabilistic density distribution \cite{li2023adaptive,li2025sparse,li2025correntropy}, consistent with prior study on generalized Bayesian framework \cite{jewson2022general}. Nonetheless, SBL-MEE suggests that such a probabilistic interpretation of the loss function in generalized Bayesian may not be necessary. In particular, MEE is essentially defined with respect to error entropy, i.e., the dispersion of error distribution without relevance to any explicit parametric distribution. Thus, this observation would motivate future research for generalized Bayesian framework to explore a broader class of sophisticated loss functions applicable to various machine learning contexts, even though such loss functions do not disclose any probability interpretation.

Despite the promising capability of our proposed SBL-MEE approach in noisy high-dimensional brain decoding, we further provide discussions regarding the limitations of SBL-MEE and potential future works. First, the kernel bandwidth $\sigma$ represents an important hyperparameter for our proposed SBL-MEE. This paper selected the optimal value for $\sigma$ by using cross validation in regression and employed a fixed value for $\sigma$ in classification. These two methods are relatively time-consuming, or probably lead to a suboptimal bandwidth. Our future studies will explore a better approach for determining $\sigma$ using a data-driven manner that could produce a proper bandwidth efficiently. In particular, this direction can be largely inspired by our previous work \cite{li2025correntropy} which proposed a score matching-based approach for selecting the bandwidth of SBCL from the residuals for an unsupervised scenario.

In addition, the SBL-MEE algorithm exhibits relatively high computational complexity. This problem primarily results from two aspects. First, to formulate the distribution $q_{\textbf{w}}(\textbf{w})$ adopting Laplacian approximation into the Gaussian distribution in (\ref{Eq.LogQwLap2}), we need to calculate the inversion of Hessian matrix $\textbf{H}\left(\textbf{w}^*\right)^{-1}$, which may be computationally prohibitive in high-dimensional task. To be specific, this step has an $\mathcal{O}\left(D^3\right)$ complexity, hence requiring a considerable computational resource. However, this problem is commonly observed for other algorithms, including SBL (on classification) \cite{yamashita2008sparse} and SBCL \cite{li2023adaptive,li2023correntropy,li2025sparse}, because they widely leverage likelihood functions non-conjugate to the prior distribution. Remarkably, another equation that involves matrix inversion is (S20) (see supplementary material) which presents in essence an iteratively reweighted least square (IRLS) update for the model parameter. While this IRLS has $\mathcal{O}\left(ND^2+D^3\right)$ complexity, it commonly converges quickly and is widely used by each method evaluated in this paper for regression contexts. Second, compared with the SBCL method that utilizes a single Gaussian kernel in the likelihood, the complexity of SBL-MEE is roughly $M$ times that of SBCL, because it adopts $M$ kernels in (\ref{Eq.AllMee}). In classification, we adopt $M=3$, while in regression, it relies on the quantization threshold $\varepsilon$, as shown in Algorithm \ref{Algo.Quantization}. In our regression experiments, the maximal value for $M$ was set to 20, which produced desirable performance. However, the future study is expected to explore the strategy for determining $M$ in regression with superior trade-off between computational cost and decoding efficacy. Besides quantization-based method in QMEE, the stochastic optimizations have also been explored as an alternative way to lighten the computational cost of MEE \cite{erdogmus2002beyond}. One possible future work is to investigate if the stochastic optimization can be combined with the quantization in QMEE, further improving the computational efficiency of the proposed SBL-MEE.

\section{Conclusion}
\label{sec:conclusion}

In this paper, we proposed a robust SBL framework by using the MEE criterion to improve the model performance for noisy and high-dimensional brain signal decoding. To be specific, we employed MEE to derive a generalized posterior that integrates the hierarchical ARD prior distributions. The proposed method was systematically evaluated on two real-world brain decoding scenarios in regression and classification contexts, respectively. The experimental results demonstrated that our proposed SBL-MEE approach not only ameliorates the decoding performance on real-world brain recording, but also facilitates the extraction of accurate physiological pattern by the parameter distribution. We also provided discussions on potential directions for further investigations.

\bibliographystyle{IEEEtran}
\bibliography{BiBib}

% Generated by IEEEtran.bst, version: 1.14 (2015/08/26)
\begin{thebibliography}{10}
\providecommand{\url}[1]{#1}
\csname url@samestyle\endcsname
\providecommand{\newblock}{\relax}
\providecommand{\bibinfo}[2]{#2}
\providecommand{\BIBentrySTDinterwordspacing}{\spaceskip=0pt\relax}
\providecommand{\BIBentryALTinterwordstretchfactor}{4}
\providecommand{\BIBentryALTinterwordspacing}{\spaceskip=\fontdimen2\font plus
\BIBentryALTinterwordstretchfactor\fontdimen3\font minus
  \fontdimen4\font\relax}
\providecommand{\BIBforeignlanguage}[2]{{%
\expandafter\ifx\csname l@#1\endcsname\relax
\typeout{** WARNING: IEEEtran.bst: No hyphenation pattern has been}%
\typeout{** loaded for the language `#1'. Using the pattern for}%
\typeout{** the default language instead.}%
\else
\language=\csname l@#1\endcsname
\fi
#2}}
\providecommand{\BIBdecl}{\relax}
\BIBdecl

\bibitem{wolpaw2000brain}
J.~Wolpaw \emph{et~al.}, ``Brain-computer interface technology: a review of the
  first international meeting,'' \emph{IEEE Transactions on Rehabilitation
  Engineering}, vol.~8, no.~2, pp. 164--173, 2000.

\bibitem{edelman2025non}
B.~J. Edelman \emph{et~al.}, ``Non-invasive brain-computer interfaces: State of
  the art and trends,'' \emph{IEEE Reviews in Biomedical Engineering}, vol.~18,
  pp. 26--49, 2025.

\bibitem{haxby2014decoding}
J.~V. Haxby, A.~C. Connolly, and J.~S. Guntupalli, ``Decoding neural
  representational spaces using multivariate pattern analysis,'' \emph{Annual
  Review of Neuroscience}, vol.~37, no.~1, pp. 435--456, 2014.

\bibitem{rybavr2022neural}
M.~Ryb{\'a}{\v{r}} and I.~Daly, ``Neural decoding of semantic concepts: A
  systematic literature review,'' \emph{Journal of Neural Engineering},
  vol.~19, no.~2, p. 021002, 2022.

\bibitem{robinson2023visual}
A.~K. Robinson, G.~L. Quek, and T.~A. Carlson, ``Visual representations:
  Insights from neural decoding,'' \emph{Annual Review of Vision Science},
  vol.~9, no.~1, pp. 313--335, 2023.

\bibitem{van2012brain}
D.~Van De~Ville and S.-W. Lee, ``Brain decoding: Opportunities and challenges
  for pattern recognition,'' \emph{Pattern Recognition}, vol.~45, no.~6, pp.
  2033--2034, 2012.

\bibitem{tang2021dimensionality}
Y.~Tang, D.~Chen, and X.~Li, ``Dimensionality reduction methods for brain
  imaging data analysis,'' \emph{ACM Computing Surveys}, vol.~54, no.~4, pp.
  1--36, 2021.

\bibitem{ball2009signal}
T.~Ball \emph{et~al.}, ``Signal quality of simultaneously recorded invasive and
  non-invasive eeg,'' \emph{NeuroImage}, vol.~46, no.~3, pp. 708--716, 2009.

\bibitem{liu2016noise}
T.~T. Liu, ``Noise contributions to the fmri signal: An overview,''
  \emph{NeuroImage}, vol. 143, pp. 141--151, 2016.

\bibitem{faul2001analysis}
A.~Faul and M.~Tipping, ``Analysis of sparse bayesian learning,''
  \emph{Advances in Neural Information Processing Systems}, vol.~14, pp. 1--7,
  2001.

\bibitem{tipping2001sparse}
M.~E. Tipping, ``Sparse bayesian learning and the relevance vector machine,''
  \emph{Journal of Machine Learning Research}, vol.~1, no. Jun, pp. 211--244,
  2001.

\bibitem{ganesh2008sparse}
G.~Ganesh \emph{et~al.}, ``Sparse linear regression for reconstructing muscle
  activity from human cortical fmri,'' \emph{NeuroImage}, vol.~42, no.~4, pp.
  1463--1472, 2008.

\bibitem{toda2011reconstruction}
A.~Toda \emph{et~al.}, ``Reconstruction of two-dimensional movement
  trajectories from selected magnetoencephalography cortical currents by
  combined sparse bayesian methods,'' \emph{NeuroImage}, vol.~54, no.~2, pp.
  892--905, 2011.

\bibitem{yoshimura2012reconstruction}
N.~Yoshimura \emph{et~al.}, ``Reconstruction of flexor and extensor muscle
  activities from electroencephalography cortical currents,''
  \emph{NeuroImage}, vol.~59, no.~2, pp. 1324--1337, 2012.

\bibitem{umeda2019decoding}
T.~Umeda \emph{et~al.}, ``Decoding of muscle activity from the sensorimotor
  cortex in freely behaving monkeys,'' \emph{NeuroImage}, vol. 197, pp.
  512--526, 2019.

\bibitem{wang2023sparse}
W.~Wang \emph{et~al.}, ``Sparse bayesian learning for end-to-end eeg
  decoding,'' \emph{IEEE Transactions on Pattern Analysis and Machine
  Intelligence}, vol.~45, no.~12, pp. 15\,632--15\,649, 2023.

\bibitem{miyawaki2008visual}
Y.~Miyawaki \emph{et~al.}, ``Visual image reconstruction from human brain
  activity using a combination of multiscale local image decoders,''
  \emph{Neuron}, vol.~60, no.~5, pp. 915--929, 2008.

\bibitem{shibata2011perceptual}
K.~Shibata \emph{et~al.}, ``Perceptual learning incepted by decoded fmri
  neurofeedback without stimulus presentation,'' \emph{Science}, vol. 334, no.
  6061, pp. 1413--1415, 2011.

\bibitem{yahata2016small}
N.~Yahata \emph{et~al.}, ``A small number of abnormal brain connections
  predicts adult autism spectrum disorder,'' \emph{Nature Communications},
  vol.~7, no.~1, pp. 1--12, 2016.

\bibitem{horikawa2017generic}
T.~Horikawa and Y.~Kamitani, ``Generic decoding of seen and imagined objects
  using hierarchical visual features,'' \emph{Nature Communications}, vol.~8,
  no.~1, pp. 1--15, 2017.

\bibitem{ganesh2018utilizing}
G.~Ganesh \emph{et~al.}, ``Utilizing sensory prediction errors for movement
  intention decoding: a new methodology,'' \emph{Science Advances}, vol.~4,
  no.~5, p. eaaq0183, 2018.

\bibitem{shi2021galvanic}
Y.~Shi \emph{et~al.}, ``Galvanic vestibular stimulation-based prediction error
  decoding and channel optimization,'' \emph{International Journal of Neural
  Systems}, vol.~31, no.~11, p. 2150034, 2021.

\bibitem{jung2000removing}
T.-P. Jung \emph{et~al.}, ``Removing electroencephalographic artifacts by blind
  source separation,'' \emph{Psychophysiology}, vol.~37, no.~2, pp. 163--178,
  2000.

\bibitem{escudero2007artifact}
J.~Escudero \emph{et~al.}, ``Artifact removal in magnetoencephalogram
  background activity with independent component analysis,'' \emph{IEEE
  Transactions on Biomedical Engineering}, vol.~54, no.~11, pp. 1965--1973,
  2007.

\bibitem{hamaneh2013automated}
M.~B. Hamaneh \emph{et~al.}, ``Automated removal of ekg artifact from eeg data
  using independent component analysis and continuous wavelet transformation,''
  \emph{IEEE Transactions on Biomedical Engineering}, vol.~61, no.~6, pp.
  1634--1641, 2014.

\bibitem{principe2010information}
J.~C. Principe, \emph{Information theoretic learning: Renyi's entropy and
  kernel perspectives}.\hskip 1em plus 0.5em minus 0.4em\relax Springer New
  York, NY, 2010.

\bibitem{liu2007correntropy}
W.~Liu, P.~P. Pokharel, and J.~C. Principe, ``Correntropy: Properties and
  applications in non-gaussian signal processing,'' \emph{IEEE Transactions on
  Signal Processing}, vol.~55, no.~11, pp. 5286--5298, 2007.

\bibitem{erdogmus2002error}
D.~Erdogmus and J.~C. Principe, ``An error-entropy minimization algorithm for
  supervised training of nonlinear adaptive systems,'' \emph{IEEE Transactions
  on Signal Processing}, vol.~50, no.~7, pp. 1780--1786, 2002.

\bibitem{chen2016insights}
B.~Chen \emph{et~al.}, ``Insights into the robustness of minimum error entropy
  estimation,'' \emph{IEEE Transactions on Neural Networks and Learning
  Systems}, vol.~29, no.~3, pp. 731--737, 2016.

\bibitem{chen2019effects}
B.~Chen \emph{et~al.}, ``Effects of outliers on the maximum correntropy
  estimation: A robustness analysis,'' \emph{IEEE Transactions on Systems, Man,
  and Cybernetics: Systems}, vol.~51, no.~6, pp. 4007--4012, 2019.

\bibitem{chen2018common}
B.~Chen \emph{et~al.}, ``Common spatial patterns based on the quantized minimum
  error entropy criterion,'' \emph{IEEE Transactions on Systems, Man, and
  Cybernetics: Systems}, vol.~50, no.~11, pp. 4557--4568, 2020.

\bibitem{zheng2020broad}
Y.~Zheng \emph{et~al.}, ``Broad learning system based on maximum correntropy
  criterion,'' \emph{IEEE Transactions on Neural Networks and Learning
  Systems}, vol.~32, no.~7, pp. 3083--3097, 2021.

\bibitem{zheng2020mixture}
Y.~Zheng \emph{et~al.}, ``Mixture correntropy-based kernel extreme learning
  machines,'' \emph{IEEE Transactions on Neural Networks and Learning Systems},
  vol.~33, no.~2, pp. 811--825, 2022.

\bibitem{li2021restricted}
Y.~Li \emph{et~al.}, ``Restricted minimum error entropy criterion for robust
  classification,'' \emph{IEEE Transactions on Neural Networks and Learning
  Systems}, vol.~33, no.~11, pp. 6599--6612, 2022.

\bibitem{li2023partial}
Y.~Li \emph{et~al.}, ``Partial maximum correntropy regression for robust
  electrocorticography decoding,'' \emph{Frontiers in Neuroscience}, vol.~17,
  p. 1213035, 2023.

\bibitem{zheng2023quantized}
Y.~Zheng, S.~Wang, and B.~Chen, ``Quantized minimum error entropy with fiducial
  points for robust regression,'' \emph{Neural Networks}, vol. 168, pp.
  405--418, 2023.

\bibitem{li2023adaptive}
Y.~Li \emph{et~al.}, ``Adaptive sparseness for correntropy-based robust
  regression via automatic relevance determination,'' in \emph{2023
  International Joint Conference on Neural Networks (IJCNN)}, 2023, pp. 1--8.

\bibitem{li2023correntropy}
Y.~Li \emph{et~al.}, ``Correntropy-based logistic regression with automatic
  relevance determination for robust sparse brain activity decoding,''
  \emph{IEEE Transactions on Biomedical Engineering}, vol.~70, no.~8, pp.
  2416--2429, 2023.

\bibitem{li2025sparse}
Y.~Li \emph{et~al.}, ``Sparse bayesian correntropy learning for robust muscle
  activity reconstruction from noisy brain recordings,'' \emph{Neural
  Networks}, vol. 182, p. 106899, 2025.

\bibitem{li2025correntropy}
Y.~Li \emph{et~al.}, ``Correntropy-based improper likelihood model for robust
  electrophysiological source imaging,'' \emph{IEEE Transactions on Medical
  Imaging}, vol.~44, no.~7, pp. 3076--3088, 2025.

\bibitem{nelder1972generalized}
J.~A. Nelder and R.~W. Wedderburn, ``Generalized linear models,'' \emph{Journal
  of the Royal Statistical Society Series A: Statistics in Society}, vol. 135,
  no.~3, pp. 370--384, 1972.

\bibitem{gelman1995bayesian}
A.~Gelman \emph{et~al.}, \emph{Bayesian data analysis}.\hskip 1em plus 0.5em
  minus 0.4em\relax Chapman and Hall, 1995.

\bibitem{blei2017variational}
D.~M. Blei, A.~Kucukelbir, and J.~D. McAuliffe, ``Variational inference: A
  review for statisticians,'' \emph{Journal of the American Statistical
  Association}, vol. 112, no. 518, pp. 859--877, 2017.

\bibitem{wipf2007new}
D.~Wipf and S.~Nagarajan, ``A new view of automatic relevance determination,''
  \emph{Advances in Neural Information Processing Systems}, vol.~20, pp. 1--8,
  2007.

\bibitem{tibshirani1996regression}
R.~Tibshirani, ``Regression shrinkage and selection via the lasso,''
  \emph{Journal of the Royal Statistical Society: Series B}, vol.~58, no.~1,
  pp. 267--288, 1996.

\bibitem{chen2018quantized}
B.~Chen \emph{et~al.}, ``Quantized minimum error entropy criterion,''
  \emph{IEEE Transactions on Neural Networks and Learning Systems}, vol.~30,
  no.~5, pp. 1370--1380, 2019.

\bibitem{parzen1962estimation}
E.~Parzen, ``On estimation of a probability density function and mode,''
  \emph{The Annals of Mathematical Statistics}, vol.~33, no.~3, pp. 1065--1076,
  1962.

\bibitem{de2013minimum}
J.~P.~M. De~Sa \emph{et~al.}, \emph{Minimum error entropy
  classification}.\hskip 1em plus 0.5em minus 0.4em\relax Springer Berlin,
  Heidelberg, 2013.

\bibitem{bissiri2016general}
P.~G. Bissiri, C.~C. Holmes, and S.~G. Walker, ``A general framework for
  updating belief distributions,'' \emph{Journal of the Royal Statistical
  Society Series B: Statistical Methodology}, vol.~78, no.~5, pp. 1103--1130,
  2016.

\bibitem{martin2022direct}
R.~Martin and N.~Syring, ``Direct gibbs posterior inference on risk minimizers:
  Construction, concentration, and calibration,'' in \emph{Handbook of
  Statistics}.\hskip 1em plus 0.5em minus 0.4em\relax Elsevier, 2022, vol.~47,
  pp. 1--41.

\bibitem{zhang2015convergence}
Y.~Zhang \emph{et~al.}, ``Convergence of a fixed-point minimum error entropy
  algorithm,'' \emph{Entropy}, vol.~17, no.~8, pp. 5549--5560, 2015.

\bibitem{mackay1992bayesian}
D.~J. MacKay, ``Bayesian interpolation,'' \emph{Neural Computation}, vol.~4,
  no.~3, pp. 415--447, 1992.

\bibitem{shimoda2012decoding}
K.~Shimoda \emph{et~al.}, ``Decoding continuous three-dimensional hand
  trajectories from epidural electrocorticographic signals in japanese
  macaques,'' \emph{Journal of Neural Engineering}, vol.~9, no.~3, p. 036015,
  2012.

\bibitem{jewson2022general}
J.~Jewson and D.~Rossell, ``General bayesian loss function selection and the
  use of improper models,'' \emph{Journal of the Royal Statistical Society
  Series B: Statistical Methodology}, vol.~84, no.~5, pp. 1640--1665, 2022.

\bibitem{yamashita2008sparse}
O.~Yamashita \emph{et~al.}, ``Sparse estimation automatically selects voxels
  relevant for the decoding of fmri activity patterns,'' \emph{NeuroImage},
  vol.~42, no.~4, pp. 1414--1429, 2008.

\bibitem{erdogmus2002beyond}
D.~Erdogmus, J.~C. Principe, and K.~E. Hild, ``Beyond second-order statistics
  for learning: A pairwise interaction model for entropy estimation,''
  \emph{Natural Computing}, vol.~1, no.~1, pp. 85--108, 2002.

\end{thebibliography}
\end{document}